\documentclass[letterpaper,english,11pt,aps,letter,superscriptaddress,nofootinbib,floatfix,pre]{revtex4}
\usepackage[T1]{fontenc}
\usepackage[latin1]{inputenc}
\usepackage{amsmath}
\usepackage{color}
\usepackage{graphicx}
\usepackage{amssymb}

\makeatletter

\usepackage{float}
\floatstyle{ruled}
\newfloat{algorithm}{tbp}{loa}
\floatname{algorithm}{Algorithm}
\renewcommand{\citet}[1]{\cite{#1}}

\usepackage{ae,aecompl}

\newenvironment{Algorithm}[1] {
  \refstepcounter{algorithm} 
  \vspace{1ex} \hrule \vspace{1ex} 
  \center{\textbf{Algorithm \thealgorithm : }#1}
  \vspace{1ex} \hrule \vspace{1ex}
  \small
} 
{ \normalsize \vspace{1ex} \hrule \vspace{1ex} }

\usepackage{babel}
\makeatother

\begin{document}

\title{A Thermodynamically-Consistent Non-Ideal Stochastic Hard-Sphere Fluid}

\author{Aleksandar Donev}

\affiliation{Lawrence Livermore National Laboratory, P.O.Box 808, Livermore, CA
94551-9900}

\affiliation{Center for Computational Science and Engineering, Lawrence Berkeley
National Laboratory, Berkeley, CA, 94720}

\author{Berni J. Alder}

\affiliation{Lawrence Livermore National Laboratory, P.O.Box 808, Livermore, CA
94551-9900}

\author{Alejandro L. Garcia}

\affiliation{Department of Physics, San Jose State University, San Jose, California,
95192}

\begin{abstract}
A grid-free variant of the Direct Simulation Monte Carlo (DSMC) method
is proposed, named the Isotropic DSMC (I-DSMC) method, that is suitable
for simulating dense fluid flows at molecular scales. The I-DSMC algorithm
eliminates all grid artifacts from the traditional DSMC algorithm;
it is Galilean invariant and microscopically isotropic. The stochastic
collision rules in I-DSMC are modified to yield a non-ideal structure
factor that gives consistent compressibility, as first proposed in
{[}\emph{Phys. Rev. Lett. 101:075902 (2008)}]. The resulting Stochastic
Hard Sphere Dynamics (SHSD) fluid is empirically shown to be thermodynamically
identical to a deterministic Hamiltonian system of penetrable spheres
interacting with a linear core pair potential, well-described by the
hypernetted chain (HNC) approximation. We apply a stochastic Enskog
kinetic theory to the SHSD fluid to obtain estimates for the transport
coefficients that are in excellent agreement with particle simulations
over a wide range of densities and collision rates. The fluctuating
hydrodynamic behavior of the SHSD fluid is verified by comparing its
dynamic structure factor against theory based on the Landau-Lifshitz
Navier-Stokes equations. We also study the Brownian motion of a nano-particle
suspended in an SHSD fluid and find a long-time power-law tail in
its velocity autocorrelation function consistent with hydrodynamic
theory and molecular dynamics calculations.
\end{abstract}
\maketitle
\newcommand{\Cross}[1]{\left|\boldsymbol{#1}\right|_{\times}}
\newcommand{\CrossL}[1]{\left|\boldsymbol{#1}\right|_{\times}^{L}}
\newcommand{\CrossR}[1]{\left|\boldsymbol{#1}\right|_{\times}^{R}}
\newcommand{\CrossS}[1]{\left|\boldsymbol{#1}\right|_{\boxtimes}}

\newcommand{\V}[1]{\boldsymbol{#1}}
\newcommand{\M}[1]{\boldsymbol{#1}}
\newcommand{\D}[1]{\Delta#1}
\newcommand{\grad}{\boldsymbol{\nabla}}
\newcommand{\Set}[1]{\mathbb{#1}}

\newcommand{\eij}{\left\{  i,j\right\}  }
\newcommand{\Wi}{\mbox{Wi}}

\newcommand{\modified}[1]{\textcolor{red}{#1}}
\newcommand{\deleted}[1]{\textcolor{red}{#1}}
\newcommand{\added}[1]{\textcolor{red}{#1}}

With the increased interest in nano- and micro-fluidics, it has become
necessary to develop tools for hydrodynamic calculations at the atomistic
scale \citet{Microfluidics_Review,MultiscaleMicrofluidics_Review}.
There are several issues present in microscopic flows that are difficult
to account for in models relying on the continuum Navier-Stokes equations.
Firstly, it is complicated to deal with boundaries and interfaces
in a way that consistently accounts for the bidirectional coupling
between the flow and (moving) complex surfaces or suspended particles.
Furthermore, it is not trivial to include thermal fluctuations in
Navier-Stokes solvers \citet{FluctuatingHydro_Garcia,FluctuatingHydro_Coveney,LLNS_S_k},
and in fact, most of the time the fluctuations are not included even
though they can be very important at instabilities \citet{FluidMixing_DSMC}
or in driving the dynamics of suspended objects \citet{LB_SoftMatter_Review,StochasticImmersedBoundary}.
Finally, since the grid cell sizes needed to resolve complex microscopic
flows are small, a large computational effort is needed even for continuum
solvers. An alternative is to use particle-based methods, which are
explicit and unconditionally stable and rather simple to implement.
The fluid particles are directly coupled to the microgeometry, for
example, they directly interact with the beads of a polymer chain.
Fluctuations occur naturally and the algorithm may be designed to
give the correct spatio-temporal correlations.

Several particle methods have been described in the literature. The
most accurate but also most expensive is molecular dynamics (MD) \citet{PolymerShear_MD},
and several coarse-grained models have been developed, such as dissipative
particle dynamics (DPD) \citet{DPD_DNA} and multi-particle collision
dynamics (MPCD) \citet{DSMC_MPCD_Gompper,DSMC_MPCD_MD_Kapral}, each
of which has its own advantages and disadvantages \citet{ParticleMesoscaleHydrodynamics}.
Our method, first proposed in Ref. \citet{SHSD_PRL}, is based on
the Direct Simulation Monte Carlo (DSMC) algorithm of Bird \citet{DSMCReview_Garcia}.
The key idea behind DSMC is to replace deterministic interactions
between the particles with stochastic momentum exchange (collisions)
between nearby particles. While DSMC is usually viewed as a kinetic
Monte Carlo algorithm for solving the Boltzmann equation for a low-density
gas, it can also be viewed as an alternative to the expensive MD in
cases where an approximate (coarse-grained) treatment of the molecular
transport is appropriate. The stochastic treatment of collisions makes
the algorithm much simpler and faster than MD, while preserving the
essential ingredients of fluctuating hydrodynamics: local momentum
conservation, linear momentum exchange on length scales comparable
to the particle size, and a similar fluctuation spectrum. 

Being composed of point particles, the DSMC fluid has no internal
structure, has an ideal gas equation of state (EOS), and is thus very
compressible. As a consequence, the density fluctuations in DSMC are
significantly larger than those in realistic liquids. Furthermore,
the speed of sound is small (comparable to the average speed of the
particles) and thus subsonic (Mach number less than one) flows are
limited to relatively small Reynolds numbers%
\footnote{For a low-density gas the Reynolds number is $Re\approx M/K$, where
$M=v_{flow}/c$ is the Mach number, and the Knudsen number $K=\lambda/L$
is the ratio between the mean free path $\lambda$ and the typical
obstacle length $L$. This shows that subsonic flows can only achieve
high $Re$ flows for small Knudsen numbers, i.e., large numbers of
DSMC particles.%
}. Efforts have been undertaken to develop coarse-grained models that
have greater computational efficiency than brute-force MD and that
have a \emph{non-ideal} EOS, such as the Lattice-Boltzmann (LB) method
\citet{NonIdeal_LB}, DPD \citet{DPD_NonIdeal}, MPCD \citet{MPCD_CBA,MPCD_CBA_2}.
The Consistent Boltzmann Algorithm (CBA) \citet{DSMC_CBA,DSMC_CBATheory},
as well as algorithms based on the Enskog equation \citet{DSMC_Enskog,DSMC_Enskog_Frezzotti},
have demonstrated that DSMC fluids can have dense-fluid compressibility,
however, they did not achieve thermodynamic consistency between the
equation of state and the fluid structure.

In this paper we describe a generalization of the traditional DSMC
algorithm suitable for dense fluid flows. By a dense fluid we mean
a fluid where the mean free path is small compared to the typical
inter-atomic distance. As a first step, we introduce a grid-free \emph{Isotropic
DSMC} (I-DSMC) method that eliminates all grid artifacts from traditional
DSMC, notably the lack of Galilean invariance and non-isotropy. The
I-DSMC fluid is still an \emph{ideal} fluid just like the traditional
DSMC fluid, that is, it has an the equation of state of an ideal gas
and does not have an internal structure as do liquids. Secondly, by
biasing the collision kernel in I-DSMC to only allow stochastic collisions
between approaching particles, we obtain the \emph{Stochastic Hard-Sphere
Dynamics} (SHSD) algorithm that is thermodynamically consistent (i.e.,
the direct calculation of compressibility from density fluctuations
agrees with the density derivative of pressure). The SHSD algorithm
is related to previous algorithms for solving the Enskog kinetic equation
\citet{DSMC_Enskog_Frezzotti,DSMC_Enskog}, and can be viewed as a
more-efficient variable-diameter stochastic modification of the traditional
hard-sphere molecular dynamics \citet{EventDriven_Alder}.

In the SHSD algorithm randomly chosen pairs of approaching particles
that lie less than a given diameter of each other undergo collisions
as if they were hard spheres of diameter equal to their actual separation.
The SHSD fluid is shown to be non-ideal, with structure and equation
of state equivalent to that of a deterministic (Hamiltonian) fluid
where penetrable spheres effectively interact with a repulsive linear
core pairwise potential. We theoretically demonstrate this correspondence
at low densities. Remarkably, we numerically find that this effective
interaction potential, similar to the quadratic core potential used
in many DPD variants, is valid at all densities. Therefore, the SHSD
fluid, as DPD, is \emph{intrinsically} thermodynamically-consistent
since it satisfies the virial theorem.

The equivalence of the structure of the SHSD fluid with the linear
core fluid enables us to use the Hypernetted Chain (HNC) approximation,
as recommended in Ref. \citet{GaussianCoreHNC}, to obtain theoretical
estimates for the pair correlation and static structure factor that
are in excellent agreement with numerical results. These further enable
us to use the Enskog-like kinetic theory developed in Ref. \citet{VariableEnskog_Transport}
to obtain accurate theoretical estimates of the transport properties
of the SHSD fluid that are also shown to be in excellent agreement
with numerics even at relatively high densities. At lower densities
the HNC approximation is not necessary and explicit expressions for
the transport coefficients can be obtained similarly to what has been
done using Green-Kubo approach for other DSMC variants \citet{DSMC_CBA}
and MPCD \citet{MPCD_VACF,MPCD_CBA,ParticleMesoscaleHydrodynamics}.

We numerically demonstrate that the hydrodynamics of the SHSD fluid
is consistent with the equations of fluctuating hydrodynamics when
the appropriate equation of state is taken into account. Specifically,
we compare the measured dynamic structure factors with that obtained
from the linearized fluctuating Navier-Stokes equations. We also calculate
the velocity autocorrelation function (VACF) for a large hard spherical
bead suspended in an SHSD fluid, demonstrating the existence of long-time
tails as predicted by hydrodynamics and found in MD simulations. The
tail is found to be in quantitative agreement with theory at lower
densities, but a discrepancy is found at higher densities, possibly
due to the strong structuring of the dense SHSD fluid.

We begin by introducing a grid-free variant of the DSMC algorithm
in Section \ref{sec:Isotropic-DSMC}. This Isotropic DSMC algorithm
simulates a stochastic particle system where particles closer than
a particle diameter collide with a certain rate. By biasing the collision
kernels to favor head-on collisions of particles, as in the hard-sphere
fluid, we obtain a non-ideal stochastic fluid in Section \ref{sec:Stochastic-Hard-Sphere}.
We develop an Enskog-like kinetic theory for this Stochastic Hard
Sphere Dynamics (SHSD) system in Section \ref{sub:Enskog-Kinetic-Theory},
which requires as input the pair correlation function. In Section
\ref{sub:Pair-Correlation-Function} we discover that the SHSD fluid
is thermodynamically consistent with a fluid of penetrable linear
core spheres, and use this equivalence to compute the pair correlation
function of the SHSD fluid using the HNC approximation. In Section
\ref{sec:Results} we show several numerical results, including a
comparison with theory for the transport coefficients and for the
dynamic structure factor, as well as a study of the hydrodynamic tails
in the velocity autocorrelation of a bead suspended in an SHSD fluid.

\section{\label{sec:Isotropic-DSMC}Isotropic DSMC}

The traditional DSMC algorithm \citet{DSMC_Bird,DSMCReview_Garcia}
starts with a time step where particles are propagated advectively,
$\V{r}_{i}^{'}=\V{r}_{i}+\V{v}_{i}\D{t}$, and sorted into a grid
of cells. Then, for each cell $c$ a certain number $N_{coll}\sim\Gamma_{sc}N_{c}(N_{c}-1)\D{t}$
of \emph{stochastic collisions} are executed between pairs of particles
randomly chosen from the $N_{c}$ particles inside the cell, where
the collision rate $\Gamma_{sc}$ is chosen based on kinetic theory.
The conservative stochastic collisions exchange momentum and energy
between two particles $i$ and $j$ that is not correlated with the
actual positions of the particles. Typically the probability of collision
is made proportional to the magnitude of the relative velocity $v_{r}=\left|\V{v}_{ij}\right|$
by using a conventional rejection procedure.

Traditional DSMC suffers from several grid artifacts, which become
pronounced when the mean free path becomes comparable to the DSMC
cell size. Firstly, the method is not Galilean invariant unless the
grid of cells is shifted randomly before each collision step, as typically
done in the MPCD algorithm \citet{DSMC_MPCD_Gompper,DSMC_MPCD_MD_Kapral}
for the same reason. This shifting is trivial in a purely particle
simulation with periodic boundary conditions, but it causes implementation
difficulties when boundaries are present and also in particle-continuum
hybrids \citet{DSMC_Hybrid}. Furthermore, traditional DSMC, unlike
MD, is not microscopically isotropic and does not conserve angular
momentum, leading to an anisotropic collisional stress tensor. Instead
of trying to work around these grid artifacts, as done for non-ideal
MPCD in Refs. \citet{MPCD_CBA,MPCD_CBA_2}, we have chosen to modify
the traditional DSMC algorithm to make the dynamics grid-free.

To ensure isotropy, all particle pairs within a collision diameter
$D$ (i.e., \emph{overlapping} particles if we consider the particles
to be spheres of diameter $D$) are considered as potential collision
partners even if they are in neighboring cells. In this way, the grid
is only used as a tool to find neighboring particles efficiently,
but does not otherwise affect the properties of the resulting stochastic
fluid. Such a grid-free DSMC variant, which we will call the Isotropic
Direct Simulation Monte Carlo (I-DSMC) method, is suitable for hydrodynamics
of dense fluids, where the mean free path is comparable or even smaller
than $D$, unlike the original DSMC which targets the dilute limit.
It is important to point out, however, that the I-DSMC is not meant
to be a replacement for traditional DSMC for rarified gas flows. In
particular, the computational efficiency is reduced by a factor of
$2-4$ over traditional DSMC due to the need to search neighboring
cells for collision partners in addition to the current cell. This
added cost is not justified at low densities, where the grid artifacts
of traditional DSMC are small. Furthermore, the I-DSMC method is not
intended as a solver for the Boltzmann equation, which was the primary
purpose of traditional DSMC \citet{DSMC_BoltzmannProof,DSMC_Granular2}.
Rather, in the limit of small time steps, the I-DSMC method simulates
the following \emph{stochastic particle system}: Particles move ballistically
in-between collisions. While two particles $i$ and $j$ are less
than a diameter $D$ apart, $r_{ij}\leq D$, there is a probability
rate $\chi D^{-1}K_{c}(\V{v}_{ij},\V{r}_{ij})$ for them to collide
and change velocities without changing their positions, where $K_{c}$
is some function of the relative position and velocity of the pair,
and the dimensionless \emph{cross-section factor} $\chi$ sets the
collisional frequency. Because the particles are penetrable, $D$
and $\chi$ may be interpreted as the range and strength, respectively,
of the interaction potential. After the collision, the pair center-of-mass
velocity does not change, ensuring momentum conservation, while the
relative velocity is drawn from a probability density $P_{c}(\V{v}_{ij}^{'};\V{v}_{ij},\V{r}_{ij})$,
such that $\left\Vert \V{v}_{ij}^{'}\right\Vert =\left\Vert \V{v}_{ij}\right\Vert $
so kinetic energy is conserved.

Once the \emph{pre- and post-collision kernels} $K_{c}$ and $P_{c}$
are specified, the properties of the resulting I-DSMC fluid are determined
by the cross-section factor $\chi$ and the \emph{density} (hard-sphere
volume fraction) $\phi=\pi ND^{3}/(6V)$, where $N$ is the total
number of particles in the simulation volume $V$. Compare this to
the deterministic hard-sphere fluid, whose properties are determined
by the volume fraction $\phi$ alone. It is convenient to normalize
the collision kernel $K_{c}$ so that for an ideal gas with a Maxwell-Boltzmann
velocity distribution the average collisional rate would be $\chi$
times larger than that of a gas of hard spheres of diameter $D$ at
low densities, $\phi\ll1$. Two particular choices for the pre-collision
kernel $K_{c}$ that we use in practice are:

\begin{description}
\item [{Traditional}] DSMC collisions (Traditional I-DSMC ideal fluid),
for which the probability of collision is made proportional to the
magnitude of the relative velocity $v_{rel}=\left\Vert \V{v}_{ij}\right\Vert $,
$K_{c}=3v_{rel}/4$. We use this kernel mainly for comparison with
traditional DSMC.
\item [{Maxwell}] collisions (Maxwell I-DSMC ideal fluid), for which $K_{c}=3\overline{v}_{rel}/4=3\sqrt{k_{B}T_{0}/\pi m}$,
where $\overline{v}_{rel}$ is the average relative velocity at equilibrium
temperature $T_{0}$. Since $K_{c}$ is a constant, all pairs collide
at the same rate, independent of their relative velocity. This kernel
is not realistic and may lead to unphysical results in cases where
there are large density and temperature gradients, however, it is
computationally most efficient since there is no rejection based on
relative velocity. We therefore prefer this kernel for problems where
the temperature dependence of the transport properties is not important,
and what we will typically mean when we say I-DSMC without further
qualification.
\end{description}
Other collision kernels may be used in I-DSMC, though we will not
consider them here \citet{DSMC_Bird}. We typically chose the traditional
DSMC post-collisional kernel $P_{c}$ in which the direction of the
post-collisional relative velocity is randomized so as to mimic the
average distribution of collision impact parameters in a low-density
hard-sphere gas. Specifically, in three dimensions the relative velocity
is rotated uniformly independent of $\V{r}_{ij}$ \citet{DSMCReview_Garcia}.
If one wishes to microscopically conserve angular momentum in I-DSMC
then the post-collisional kernel has to use the actual positions of
the colliding particles. Specifically, the component of the relative
velocity perpendicular to the line joining the colliding particles
should remain unchanged, while the parallel component should be reversed. 

Note that a pairwise Anderson thermostat proposed within the context
of MD/DPD by Lowe \citet{DPDSchmidtNumbers} adds I-DSMC-like collisions
to ordinary MD. In addition to algorithmic differences with I-DSMC,
in Lowe's method the post-collisional kernel is such that it preserves
the normal component of the relative velocity (thus conserving angular
momentum), while the parallel component is thermalized by drawing
from a Maxwell-Boltzmann distribution. We strive to preserve exact
conservation of both momentum and energy in the collision kernels
we use, without artificial energy transport via thermostating.

With a finite time step, the I-DSMC method can be viewed as a time-driven
kinetic Monte Carlo algorithm to solve the Master Equation for the
stochastic particle system described above. Unlike the singular kernel
in the Boltzmann equation, this Master Equation has a mollified collision
kernel with a finite compact support $D$ \citet{StochasticFluid_Euler,VariableEnskog_Transport}.
The traditional DSMC method also mollifies the collision kernel by
considering particles within the same collision cell, of size $L_{c}$,
as possible collision partners. This DSMC cell size is much larger
than a molecular diameter, $D_{m}$, in fact, for low densities it
is a fraction (typically a quarter) of the mean free path. The molecular
properties enter in traditional DSMC only in the form of collisional
cross-sections $\sigma\sim D_{m}^{2}$. In light of this, for rarified
gas flows, the collision diameter $D$ in I-DSMC should be considered
the equivalent of the cell length $L_{c}$, and \emph{not} $D_{m}$.
Traditional DSMC is designed to reproduce a collision rate per particle
per unit time equal to the Boltzmann rate, $\Gamma_{B}(D_{m})=CD_{m}^{2}$,
where $C$ is a constant. The I-DSMC method is designed to reproduce
a collision rate\[
\Gamma_{I-DSMC}=\chi\Gamma_{B}(D)=\chi CD^{2},\]
and therefore by choosing\[
\chi=\chi_{B}=\left(\frac{D_{m}}{D}\right)^{2}\]
we get $\Gamma_{I-DSMC}=\Gamma_{B}(D_{m})$. Therefore, if I-DSMC
is used to simulate the transport in a low-density gas of hard-sphere
of diameter $D_{m}$, the collision diameter $D$ should be chosen
to be some fraction of the mean free path $\lambda$ (say, $D\approx\lambda/4\gg D_{m}$),
and the cross-section factor set to $\chi_{B}\sim(D_{m}/\lambda)^{2}\ll1$.
At higher densities $\chi_{B}$ starts becoming comparable to unity
and thus it is no longer possible to separate the kinetic and collisional
time scales as assumed in traditional DSMC. Note that I-DSMC is designed
for dense fluids so while it is possible to apply it in simulating
rarefied gases it will not be as computationally efficient as traditional
DSMC.

\subsection{Performing Stochastic Collisions}

In I-DSMC, stochastic collisions are processed at the begining of
every time step of duration $\D{t}$, and then each particle $i$
is streamed advectively with constant velocity $\V{v}_{i}$. During
the collision step, we need to randomly and without bias choose pairs
of overlapping particles for collision, given the current configuration
of the system. This can be done, as in traditional DSMC, using a rejection
Monte Carlo technique. Specifically, we need to choose a large number
$N_{tc}^{(tot)}=\Gamma_{tc}^{(tot)}N_{pairs}\D{t}$ of \emph{trial
collision pairs}, and then accept the fraction of them that are actually
overlapping as \emph{collision candidates}. Here $N_{pairs}$ is the
number of possibly-overlapping pairs, for example, as a first guess
one can include all pairs, $N_{pairs}=N(N-1)/2$. The probability
for choosing one of the overlapping pairs as a collision candidate
is simply $\Gamma_{tc}^{(tot)}\D{t}$. If the probability of accepting
a candidate pair $ij$ for an actual collision is $p_{ij}^{(acc)}$
and $\D{t}$ is sufficiently small, then the probability \emph{rate}
to actually collide particles $i$ and $j$ while they are overlapping
approaches $\Gamma_{ij}=p_{ij}^{(acc)}\Gamma_{tc}^{(tot)}$. The goal
is to choose the trial collision frequency $\Gamma_{tc}^{(tot)}$
and $p_{ij}^{(acc)}$ such that $\Gamma_{ij}=\chi D^{-1}K_{c}(\V{v}_{ij},\V{r}_{ij})$.

The efficiency of the algorithm is increased if the probability of
accepting trial collisions is increased. In order to increase the
acceptance probability, one should reduce $N_{pairs}$ to be closer
to the number of actually overlapping pairs, ideally, one would build
a list of all the overlapping pairs (making $N_{pairs}$ linear instead
of quadratic in $N$). This is however expensive, and a reasonable
compromise is to use collision cells similarly to what is done in
classical DSMC and also MD algorithms. Namely, the spatial domain
of the simulation is divided into\emph{ cells} of length $L_{c}\gtrapprox D$,
and for each cell a linked list $\mathcal{L}_{c}$ of all the particles
in that cell is maintained. All pairs of particles that reside in
the same or neighboring cells are considered as potential collision
partners, and here we include the cell itself in its list of neighboring
cells, i.e., each cell has $3^{d}$ neighbors, where $d$ is the spatial
dimension.

To avoid any spatial correlations (inhomogeneity and non-isotropy),
trial collision pairs should be chosen at random one by one. This
would require first choosing a pair of neighboring cells with the
correct probability, and then choosing a particle from each cell (rejecting
self-collisions). This is rather expensive to do, especially at lower
$\chi$, when few actual collisions occur at each time step, and we
have therefore chosen to use a method that introduces a small bias
each time step, but is unbiased over many time steps. Specifically,
we visit the cells one by one and for each cell $c$ we perform $N_{tc}^{(c)}=\Gamma_{tc}^{(c)}N_{c}N_{p}\D{t}$
trial collisions between one of the $N_{c}$ particles in that cell
and one of the $N_{p}$ particles in the $3^{d}$ neighboring cells,
rejecting self-collisions. Here $\Gamma_{tc}^{(c)}$ is a \emph{local}
trial collision rate and it may depend on the particular cell $c$
under consideration. Note that each of the $N_{c}(N_{c}-1)$ trial
pairs $ij$ where both $i$ and $j$ are in cell $c$ is counted twice,
and similarly, any pair where $i$ and $j$ are in different cells
$c$ and $c^{\prime}$ is included as a trial pair twice, once when
each of the cells $c$ and $c^{\prime}$ is considered. Also note
that it is important not to visit the cells in a fixed order during
every time step. Unlike in traditional cells, where cells are independent
of each other and can be visited in an arbitrary order (even in parallel),
in I-DSMC it is necessary to ensure isotropy by visiting the cells
in a random order, different every time step.

For the Maxwell pre-collision kernel, once a pair of overlapping particles
$i$ and $j$ is found a collision is performed without additional
rejection, therefore, we set $\Gamma_{tc}^{(c)}=\chi D^{-1}K_{c}/2=3\chi(2D)^{-1}\sqrt{k_{B}T_{0}/\pi m}=\mbox{const}$;
note that we have divided by two because of the double counting of
each trial pair. For the traditional pre-collision kernel, and, as
we shall see shortly, the SHSD pre-collision kernel, additional rejection
based on the relative velocity $\V{v}_{ij}$ is necessary. As in the
traditional DSMC algorithm, we estimate an upper bound for the maximal
value of the pre-collision kernel $K_{c}^{(max)}$ among the pairs
under consideration and set $\Gamma_{tc}^{(c)}=\chi D^{-1}K_{c}^{(max)}/2$.
We then perform an actual collision for the trial pair $ij$ with
probability\[
p_{ij}^{c}=K_{c}(\V{v}_{ij},\V{r}_{ij})/K_{c}^{(max)},\]
giving the correct collision probability for every overlapping pair
of particles. For the traditional pre-collision kernel $K_{c}^{(max)}=3v_{rel}^{(max)}/4$,
where $v_{rel}^{(max)}$ is as tight an estimate of the maximum relative
speed as possible. In the traditional DSMC algorithm $v_{rel}^{(max)}$
is a global bound obtained by simply keeping track of the maximum
particle speed $v_{max}$ and taking $v_{rel}^{(max)}=2v_{max}$ \citet{DSMCReview_Garcia}.
In I-DSMC, we obtain a local estimate of $v_{rel}^{(max)}$ for each
cell $c$ that is visited, thus increasing the acceptance rate and
improving efficiency.

Algorithm \ref{ProcessCollisions} specifies the procedure for performing
collisions in the I-DSMC method. The algorithm is to a large degree
collision-kernel independent, and in particular, the same algorithm
is used for ideal and non-ideal stochastic fluids. As already explained,
the size of the cells should be chosen to be as close as possible
but still larger than the particle diameter $D$. The time step should
be chosen such that a typical particle travels a distance $l_{\D{t}}\approx v_{th}\D{t}\sim D\delta t$,
where the typical thermal velocity $v_{th}=\sqrt{k_{B}T_{0}/m}$ and
$\delta t$ is a dimensionless time step, which should be kept reasonably
smaller than one, for example, $\delta t\lessapprox0.25$. It is also
important to ensure that each particle does not, on average, undergo
more than one collision per time step; we usually keep the number
of collisions per particle per time step less than one half. Since
a typical value of the pre-collision kernel is $K_{c}\sim v_{th},$
the number of collisions per particle per time-step can easily be
seen to be on the order of \[
N_{cps}\sim\chi\frac{v_{th}}{D}\cdot\frac{N}{V}V_{p}\cdot\D{t}=\chi\phi\delta t,\]
where $V_{p}\sim D^{3}$ is a particle volume. Therefore, unless $\chi\phi\gg1$,
choosing a small dimensionless time step $\delta t$ will ensure that
the collisional frequency is not too large, $N_{cps}\ll1$. With these
conditions observed, we find little dependence of the fluid properties
on the actual value of $\delta t$.

\begin{Algorithm}{\label{ProcessCollisions}Processing of stochastic
collisions between overlapping particles at a time-step in the I-DSMC
method.}

\begin{enumerate}
\item Sample a random permutation of the cell numbering $\mathcal{P}$.
\item Visit the cells one by one in the random order given by $\mathcal{P}$.
For each cell $c$, do the following steps if the number of particles
in that cell $N_{c}>0$, otherwise move on to the next cell.
\item Build a list $\mathcal{L}_{1}$ of the $N_{c}$ particles in the cell
and at the same time find the largest particle speed in that cell
$v_{1}^{max}$. Also keep track of the second largest speed in that
cell $v_{2}^{max}$, which is an estimate of the largest possible
speed of a collision partner for the particle with speed $v_{1}^{max}$.
\item Build a list of the $N_{p}$ particles in the set of $3^{d}$ cells
that neighbor $c$, including the cell $c$ itself and respecting
the proper boundary conditions. Also update $v_{2}^{max}$ if any
of the potential collision partners not in cell $c$ have speeds greater
than $v_{2}^{max}$.
\item Determine the number of trial collisions between a particle in cell
$c$ and a neighboring particle by rounding to an integer \citet{DSMCReview_Garcia}
the expected value\[
N_{tc}=\Gamma_{tc}N_{c}N_{p}\D{t},\]
where $\D{t}$ is the time step. Here the local trial collision rate
is\[
\Gamma_{tc}=\frac{\chi K_{c}^{max}}{2D},\]
where $K_{c}^{max}$ is an upper bound for the pre-collision kernel
among all candidate pairs. For Maxwell collisions $K_{c}^{max}=3\sqrt{k_{B}T_{0}/\pi m}$,
and for traditional collisions $K_{c}^{max}=3v_{rel}^{(max)}/4$,
where $v_{max}^{(rel)}=(v_{1}^{max}+v_{2}^{max})$ is a local upper
bound on the relative speed of a colliding pair.
\item Perform trial collisions by randomly selecting $N_{tc}$ pairs of
particles $i\in\mathcal{L}_{1}$ and $j\in\mathcal{L}_{2}$. For each
pair, do the following steps if $i\neq j$:

\begin{enumerate}
\item Calculate the distance $l_{ij}$ between the centroids of particles
$i$ and $j$, and go to the next pair if $l_{ij}>D$.
\item Calculate the collision kernel $K_{ij}^{c}=K_{c}(\V{v}_{ij},\V{r}_{ij})$,
and go to the next pair if $K_{ij}^{c}=0$.
\item Sample a random uniform variate $0<r\leq1$ and go to the next pair
if $K_{ij}^{c}\leq rK_{c}^{max}$ (note that this step can be skipped
in Maxwell I-DSMC since $K_{ij}^{c}=K_{c}^{max}$).
\item Process a stochastic collision between the two particles by updating
the particle velocities by sampling the post-collision kernel $P_{c}(\V{v}_{ij}^{'};\V{v}_{ij},\V{r}_{ij})$.
For ideal fluids we perform the usual stochastic DSMC collision by
randomly rotating $\V{v}_{ij}$ to obtain $\V{v}_{ij}^{'}$, independent
of $\V{r}_{ij}$.
\end{enumerate}
\end{enumerate}
\end{Algorithm}

\section{\label{sec:Stochastic-Hard-Sphere}Stochastic Hard Sphere Dynamics}

The traditional DSMC fluid has no internal structure so it has an
ideal gas equation of state (EOS), $p=PV/Nk_{B}T=1$, and is thus
very compressible. As for the classical hard-sphere fluid, the pressure
of fluids with stochastic collisions consists of two parts, the usual
\emph{kinetic} contribution that gives the ideal-gas pressure $p_{k}=1$,
and a \emph{collisional} contribution proportional to the virial $p_{c}\sim\left\langle \left(\V{v}_{ij}\cdot\V{r}_{ij}\right)^{\prime}-\left(\V{v}_{ij}\cdot\V{r}_{ij}\right)\right\rangle _{c}$,
where the average is over stochastic collisions and primes denote
post-collisional values. The virial vanishes for collision kernels
where velocity updates and positions are uncorrelated, as in traditional
DSMC, leaving only the ideal-gas kinetic contribution. In order to
introduce a non-trivial equation of state it is necessary to either
give an additional displacement $\D{\V{r}_{ij}}$ to the particles
that is parallel to $\V{v}_{ij}$, or to bias the momentum exchange
$\D{\V{p}_{ij}}=m\D{\V{v}_{ij}}$ to be (statistically) aligned to
$\V{r}_{ij}$. The former approach has already been investigated in
the Consistent Boltzmann Algorithm (CBA) \citet{DSMC_CBA,DSMC_CBATheory}.
This algorithm was named {}``consistent'' because both the transport
coefficients and the equation of state are consistent with those of
a hard-sphere fluid to lowest order in density, unlike traditional
DSMC which only matches the transport coefficients. However, CBA is
not \emph{thermodynamically} consistent since it modifies the compressibility
without affecting the density fluctuations (i.e., the structure of
the fluid is still that of a perfect gas).

Here we explore the option of biasing the stochastic momentum exchange
based on the position of the colliding particles. What we are trying
to emulate through this bias is an effective repulsion between overlapping
particles. This repulsion will be maximized if we make $\D{\V{p}_{ij}}$
parallel to $\V{r}_{ij}$, that is, if we use the hard-sphere collision
rule $P_{c}(\V{v}_{ij}^{'};\V{v}_{ij},\V{r}_{ij})=\delta(\V{v}_{ij}+2v_{n}\hat{\V{r}}_{ij})$,
where $v_{n}=-\V{v}_{ij}\cdot\hat{\V{r}}_{ij}$ is the normal component
of the relative velocity. Explicitly, we collide particles as if they
are elastic hard spheres of diameter equal to the distance between
them at the time of the collision,\begin{align}
\V{v}_{i}^{\prime}= & \V{v}_{i}+v_{n}\hat{\V{r}}_{ij}\nonumber \\
\V{v}_{j}^{\prime}= & \V{v}_{j}-v_{n}\hat{\V{r}}_{ij}.\label{eq:P_c_hard_spheres}\end{align}
Such collisions produce a positive virial only if the particles are
approaching each other, i.e., if $v_{n}>0$, therefore, we reject
collisions among particles that are moving apart, $K_{c}(\V{v}_{ij},\V{r}_{ij})\sim\Theta(v_{n})$,
where $\Theta$ denotes the Heaviside function. Note that the hard-sphere
post-collision rule (\ref{eq:P_c_hard_spheres}) strictly conserves
angular momentum in addition to linear momentum and energy and can
be used with other pre-collision kernels (e.g., Maxwell) if one wishes
to conserve angular momentum.

To avoid rejection of candidate collision pairs and thus make the
algorithm most efficient, it would be best if the pre-collision kernel
$K_{c}$ is independent of the relative velocity as for Maxwell collisions.
However, without rejection based on the normal $v_{n}$ or relative
$v_{r}$ speeds, fluctuations of the local temperature $T_{c}$ would
not be consistently coupled to the local pressure. Namely, without
rejection the \emph{local} collisional frequency $\Gamma_{sc}$ would
be independent of $T_{c}$ and thus the collisional contribution to
the pressure $p_{c}\sim\left\langle \D{\V{v}_{ij}}\cdot\V{r}_{ij}\right\rangle _{c}\sim\Gamma_{sc}\sqrt{T_{c}}$
would be $p_{c}\sim\sqrt{T_{c}}$ instead of $p_{c}\sim T_{c}$, as
is required for a fluid with no internal energy \citet{MPCD_CBA,MPCD_CBA_2}.
Instead, as for hard spheres, we require that $\Gamma_{sc}\sim\sqrt{T_{c}}$,
which is satisfied if the collision kernel is linear in the magnitude
of the relative velocity. For DSMC the collisional rules can be manipulated
arbitrarily to obtain the desired transport coefficients, however,
for non-ideal fluids thermodynamic requirements eliminate some of
the freedom. This important observation has not been taken into account
in other algorithms that randomize hard-sphere molecular dynamics
\citet{PPM_DSMC}, but has been used in the non-ideal MPCD algorithm
in order to obtain thermodynamic consistency \citet{MPCD_CBA,MPCD_CBA_2}.

There are two obvious choices for a pre-collision kernel that are
linear in the magnitude of the relative velocity. One is to use the
relative speed, $K_{c}\sim v_{r}$, as in the traditional DSMC algorithm,
and the other is to use the hard-sphere pre-collision kernel, $K_{c}\sim v_{n}$.
We have chosen to make the collision probability linear in the normal
speed $v_{n}$, specifically, we take $K_{c}=3v_{n}\Theta(v_{n})$
to define the \emph{Stochastic Hard-Sphere Dynamics} (SHSD) fluid,
similarly to what has previously been done in the Enskog DSMC algorithm
\citet{DSMC_Enskog,DSMC_Enskog_Frezzotti} and in non-ideal MPCD \citet{MPCD_CBA,MPCD_CBA_2}.
These choices for the collision kernels make the SHSD fluid identical
to the one proposed in Ref. \citet{StochasticFluid_Euler} for the
purposes of proving convergence of a microscopic model to the Navier-Stokes
equations. Specifically, the singular Boltzmann hard-sphere collision
kernel is mollified in Ref. \citet{StochasticFluid_Euler} to obtain
the SHSD collision kernel and then the low-density hydrodynamic limit
is considered. 

The non-ideal SHSD fluid is simulated by the I-DSMC method, in the
limit of sufficiently small time steps. However, it is important to
observe that the SHSD fluid is defined independently of any temporal
discretization used in computer simulations, just like a Hamiltonian
fluid is defined through the equations of motion independently of
Molecular Dynamics (MD). To summarize, in the SHSD algorithm we use
the following collision kernels in Algorithm \ref{ProcessCollisions}:\begin{align*}
K_{c}= & 3v_{n}\Theta(v_{n})\mbox{ and }K_{c}^{(max)}=3v_{max}^{(rel)}\\
P_{c}(\V{v}_{ij}^{\prime})= & \delta(\V{v}_{ij}+2v_{n}\hat{\V{r}}_{ij})\\
\mbox{where }v_{n} & =-\V{v}_{ij}\cdot\hat{\V{r}}_{ij}.\end{align*}
Note that considering particles in neighboring cells as collision
partners is essential in SHSD in order to ensure isotropy of the collisional
(non-ideal) component of the pressure tensor. It is also important
to traverse the cells in random order when processing collisions,
as well as to ensure a sufficiently small time step is used to faithfully
simulate the SHSD fluid. Note that the SHSD algorithm strictly conserves
both momentum and energy independent of the time step.

\subsection{\label{sub:Enskog-Kinetic-Theory}Enskog Kinetic Theory}

In this section we develop some kinetic equations for the SHSD fluid
that are inspired by the Enskog theory of hard-sphere fluids. Remarkably,
it turns out that these sorts of kinetic equations have already been
studied in the literature for purely theoretical purposes.

\subsubsection{BBGKY Hierarchy}

The full Bogoliubov-Born-Green-Kirkwood-Yvon (BBGKY) hierarchy of
Master equations describing the SHSD fluid is derived in Ref. \citet{StochasticFluid_Euler}.
Specifically, the evolution of the $s$-particle distribution function
$f_{s}(t;\V{r}_{1},\V{v}_{1},\ldots,\V{r}_{s},\V{v}_{s})$ is governed
by\begin{align}
\frac{\partial f_{s}}{\partial t}+\sum_{i=1}^{s}\V{v}_{i}\cdot\grad_{\V{r}_{i}}f_{s}= & 3\chi D^{2}\int_{0}^{1}dx\int_{\mathcal{R}^{3}}d\V{v}_{j}\int_{S_{+}^{2}}d\hat{\V{r}}_{ij}\quad x^{2}\sum_{i=1}^{s}v_{n}\nonumber \\
\bigl[ & f_{s+1}(t;\V{r}_{1},\V{v}_{1},\ldots,\V{r}_{i},\V{v}_{i}',\ldots,\V{r}_{s},\V{v}_{s},\V{r}_{i}+x\hat{\V{r}}_{ij},\V{v}_{j}^{\prime})\nonumber \\
- & f_{s+1}(t;\V{r}_{1},\V{v}_{1},\ldots,\V{r}_{i},\V{v}_{i},\ldots,\V{r}_{s},\V{v}_{s},\V{r}_{i}-x\hat{\V{r}}_{ij},\V{v}_{j})\bigr]\label{BBGKY_df_dt}\end{align}
which takes into account the contribution from collisions of one of
the $s$ particles, particle $i$, with another particle $j$ that
is at a distance $r_{ij}=xD$ away, $0\leq x\leq1$. Here $S_{+}^{2}$
denotes the fraction of the unit sphere for which $v_{n}=-\hat{\V{r}}_{ij}\cdot(\V{v}_{i}-\V{v}_{j})\geq0$,
and $\V{v}_{i}'=\V{v}_{i}+v_{n}\hat{\V{r}}_{ij}$ and $\V{v}_{j}^{\prime}=\V{v}_{j}-v_{n}\hat{\V{r}}_{ij}$.
Just like the BBGKY hierarchy for Hamiltonian fluids, Eqs. (\ref{BBGKY_df_dt})
are exact, however, they form an infinite unclosed system in which
the $(s+1)$-particle distribution function appears in the equation
for the $s$-particle distribution function. As usual, we need to
make an anzatz to truncate and close the system, as we do next.

\subsubsection{\label{SectionEnskogEq}Thermodynamic and Transport Properties}

The hydrodynamics of the SHSD fluid is well-described by a kinetic
equation for the single-particle probability distribution $f(t,\V{r},\V{v})\equiv f_{1}(t;\V{r},\V{v})$
obtained by making the common molecular chaos assumption about the
two-particle distribution function,\[
f_{2}(t;\V{r}_{1},\V{v}_{1},\V{r}_{2},\V{v}_{2})=g_{2}(\V{r}_{1},\V{r}_{2};n)f(t,\V{r}_{1},\V{v}_{1})f(t,\V{r}_{2},\V{v}_{2}),\]
where $g_{2}(\V{r}_{i},\V{r}_{j};n)$ is the non-equilibrium pair
distribution function that is a functional of the local number density
$n(\V{r})$. At global equilibrium $n(\V{r})=\mbox{const}$ and $g_{2}\equiv g_{2}(r_{ij})$
depends only on the radial distance once the equilibrium density $n$
and cross-section factor $\chi$ are specified. Substituting the above
assumption for $f_{2}$ in the first equation of the BBGKY hierarchy
(\ref{BBGKY_df_dt}), we get a stochastic revised Enskog equation
of the form studied in Ref. \citet{VariableEnskog_Transport},

\begin{align}
\frac{\partial f(t,\V{r},\V{v})}{\partial t}+\V{v}\cdot\grad_{\V{r}}f(t,\V{r},\V{v})= & 3\chi D^{2}\int_{0}^{1}dx\int_{\mathcal{R}^{3}}d\V{w}\int_{S_{+}^{2}}d\V{e}\quad x^{2}v_{n}\nonumber \\
\bigl[ & g_{2}(\V{r},\V{r}+x\V{e};n)f(t,\V{r},\V{v}')f(t,\V{r}+x\V{e},\V{w}')\nonumber \\
- & g_{2}(\V{r},\V{r}-x\V{e};n)f(t,\V{r},\V{v})f(t,\V{r}-x\V{e},\V{w})\bigr]\label{Enskog_df_dt}\end{align}
where $v_{n}=-\V{e}\cdot(\V{v}-\V{w})\geq0$, $\V{v}'=\V{v}+\V{e}v_{n}$
and $\V{w}'=\V{w}-\V{e}v_{n}$.

The standard second-order Chapman-Enskog expansion has been carried
out for the {}``stochastic Enskog'' equation of the same form as
Eq. (\ref{Enskog_df_dt}) in Ref. \citet{VariableEnskog_Transport},
giving the equation of state (EOS) $p=PV/Nk_{B}T$, and estimates
of the diffusion coefficient $\zeta$, the shear $\eta$ and bulk
$\eta_{B}$ viscosities, and thermal conductivity $\kappa$ of the
SHSD fluid. The expressions in Ref. \citet{VariableEnskog_Transport}
ultimately express the transport coefficients in terms of various
dimensionless integer moments of the pair correlation function $g_{2}(x=r/D)$,
$x_{k}=\int_{0}^{1}x^{k}g_{2}(x)dx$, specifically, \begin{align}
p-1= & 12\phi\chi x_{3},\label{SHSD_p}\\
\zeta/\zeta_{0}= & \frac{\sqrt{\pi}}{48\phi\chi x_{2}},\\
\eta_{B}/\eta_{0}= & \frac{48\phi^{2}\chi x_{4}}{\pi^{3/2}},\\
\eta/\eta_{0}= & \frac{5}{48\sqrt{\pi}\chi x_{2}}(1+\frac{24\phi\chi x_{3}}{5})^{2}+\frac{3}{5}\eta_{B},\mbox{ and}\label{SHSD_eta}\\
\kappa/\kappa_{0}= & \frac{25}{64\sqrt{\pi}\chi x_{2}}(1+\frac{36\phi\chi x_{3}}{5})^{2}+\frac{3}{2}\eta_{B},\end{align}
where $\zeta_{0}=D\sqrt{k_{B}T/m}$, $\eta_{0}=D^{-2}\sqrt{mk_{B}T}$
and $\kappa_{0}=k_{B}D^{-2}\sqrt{k_{B}T/m}$ are natural units. These
equations are very similar to the ones in the Enskog theory of the
hard-sphere fluid except that various coefficients are replaced with
moments of $g_{2}(x)$. In order to use these equations, however,
we need to have a good approximation to the pair correlation function,
i.e., to the structure of the SHSD fluid. It is important to point
out that Eq. (\ref{SHSD_p}) is \emph{exact} as it can be derived
directly from the definition of the collisional contribution to the
pressure.

\subsection{\label{sub:Pair-Correlation-Function}Pair Correlation Function}

In this section we study the structure of the SHSD fluid, theoretically
at low densities, and then numerically at higher densities. We find,
surprisingly, that there is a thermodynamic correspondence between
the stochastic SHSD fluid and a deterministic penetrable-sphere fluid.

\subsubsection{Low Densities}

In order to understand properties of the SHSD fluid as a function
of the density $\phi$ and the cross-section factor $\chi$, we first
consider the equilibrium pair correlation function $g_{2}(r)$ at
low densities, where correlations higher than pairwise can be ignored.
We consider the cloud of point walkers $ij$ representing the $N(N-1)/2$
pairs of particles, each at position $\V{r}=\V{r}_{i}-\V{r}_{j}$
and with velocity $\V{v}=\V{v}_{i}-\V{v}_{j}$. If one of these walkers
is closer than $D$ to the origin, $r\leq D$, and is approaching
the origin, $v_{n}>0$, it reverses its radial speed as a stochastic
process with a time-dependent rate\emph{ $\Gamma=\left|v_{n}\right|\Gamma_{0}$},
where $\Gamma_{0}=3\chi/D$ is the collision frequency. A given walker
corresponding to pair $ij$ also undergoes stochastic spatially-unbiased
velocity changes with some rate due to the collisions of $i$ with
other particles. At low densities we can assume that these additional
collisions merely thermalize the velocities to a Maxwell-Boltzmann
distribution but not otherwise couple with the radial dependence of
the one-particle distribution function $f_{pairs}(\V{v},\V{r})$ of
the $N(N-1)/2$ walkers. Inside the core $r\leq D$ this distribution
of pair walkers satisfies a kinetic equation\begin{equation}
\frac{\partial f_{pairs}}{\partial t}-v_{n}\frac{\partial f_{pairs}}{\partial r}=\left\{ \begin{array}{c}
-\Gamma f_{pairs}\mbox{ if }v_{n}\geq0\\
\Gamma f_{pairs}\mbox{ if }v_{n}<0\end{array}\right.=-\Gamma_{0}v_{n}f_{pairs},\label{eq:df_dt_low_density}\end{equation}
where the term $-\Gamma f_{pairs}$ is a loss term for approaching
pairs due to their collisions, while $\Gamma f_{pairs}$ is a gain
term for pairs that are moving part due to collisions of approaching
pairs that then reverse their radial speed. At equilibrium, $\partial f_{pairs}/\partial t=0$
and $v_{n}$ cancels on both sides, consistent with choosing collision
probability linear in $\left|v_{n}\right|$, giving $\partial f_{pairs}/\partial r=3\chi D^{-1}f_{pairs}$.
At equilibrium, the distribution of the point walkers in phase space
ought to be of the separable form \emph{$f_{pairs}(\V{v},\V{r})=f_{pairs}(v_{n},r)\sim g_{2}(r)\exp(-mv_{n}^{2}/4kT)$,}
giving $dg_{2}(r)/dr=3\chi D^{-1}g_{2}(r)$ for $r\leq D$ and zero
otherwise, with solution \begin{equation}
g_{2}(x=r/D)=\left\{ \begin{array}{c}
\exp\left[3\chi(x-1)\right]\mbox{ for }x\leq1\\
1\mbox{ for }x>1\end{array}\right.\label{SHSD_g2_low}\end{equation}

Indeed, numerical experiments confirmed that at sufficiently low densities
the equilibrium $g_{2}$ for the SHSD fluid has the exponential form
(\ref{SHSD_g2_low}) inside the collision core. From statistical mechanics
we know that for a deterministic Hamiltonian particle system with
a pairwise potential $U(r)$ at low density $g_{2}^{U}=\exp[-U(r)/kT]$.
Therefore, the low density result (\ref{SHSD_g2_low}) is consistent
with an effective \emph{linear core} pair potential\begin{equation}
U_{eff}(r)/kT=3\chi(1-x)\Theta(1-x).\label{SHSD_U_eff}\end{equation}
Note that this repulsive potential is similar to the quadratic core
potential used in DPD and strictly vanishes outside of the overlap
region, as expected. Also note that the cross-section factor $\chi$
plays the role that $U(0)/kT$ plays in the system of penetrable spheres
interacting with a linear core pairwise potential.

As pointed out earlier, Eq. (\ref{SHSD_p}) is exact. At the same
time, it is equivalent to the virial theorem for the linear core potential.
Therefore, if the pair correlation functions of the SHSD fluid and
the linear core fluid are truly identical, the pressure of the SHSD
fluid is identically equal to that of the corresponding penetrable
sphere system. As a consequence, thermodynamic consistency between
the structure {[}$g_{2}(x)$ and $S(k)$] and equation-of-state {[}$p(\phi)$]
is guaranteed to be exact for the SHSD fluid.

\subsubsection{Equivalence to the Linear Core Penetrable Sphere System}

\begin{figure}[tbph]
\begin{centering}
\includegraphics[width=0.495\textwidth]{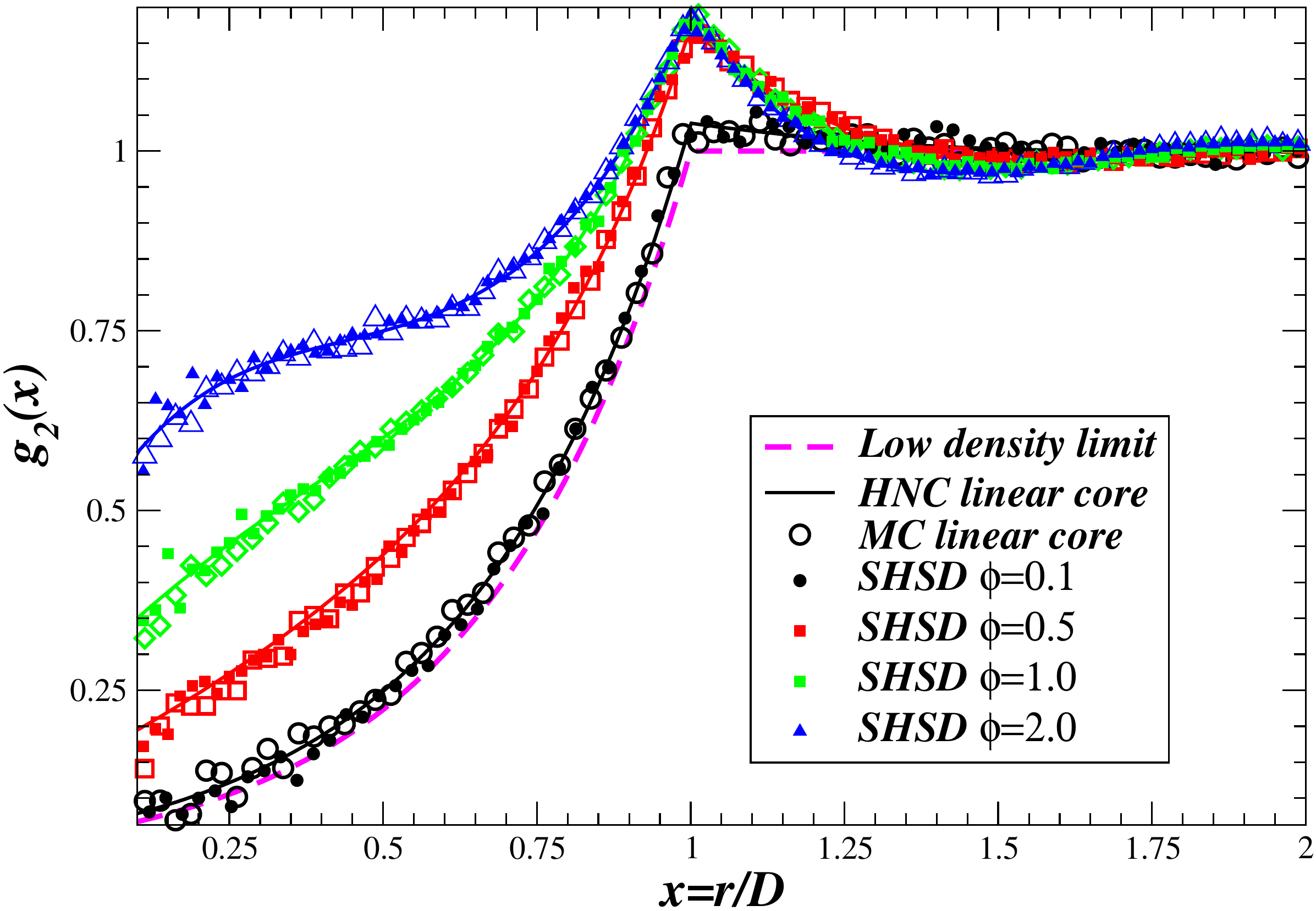}\includegraphics[width=0.495\textwidth]{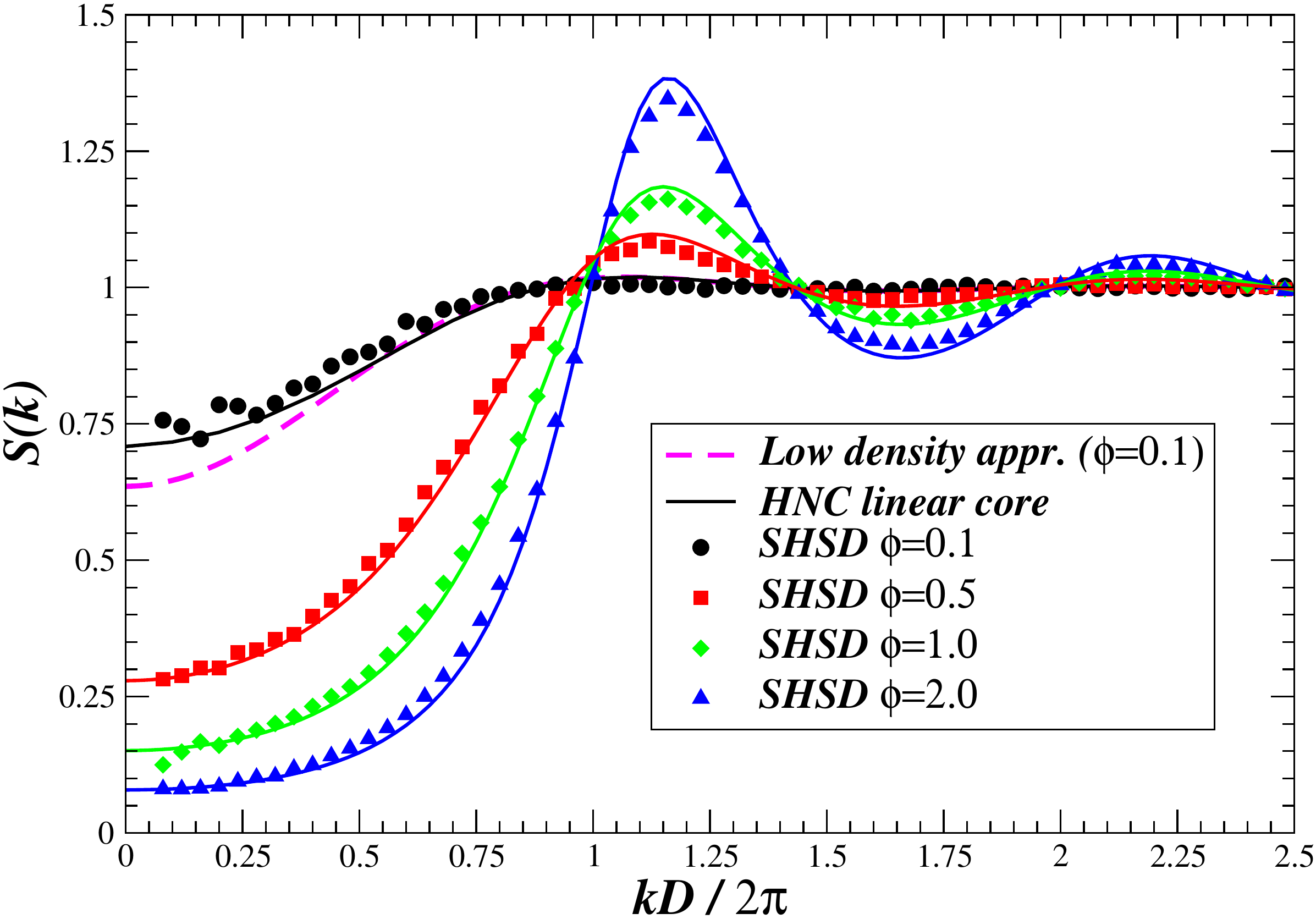}
\par\end{centering}

\caption{\label{SHSD_g2}(\emph{Left}) Equilibrium pair correlation function
of the SHSD fluid (solid symbols, $N=10^{4}$ particles in a cubic
periodic box), compared to Monte Carlo simulations (open symbols,
$N=10^{4}$ particles in a cubic periodic box) and numerical solution
of the HNC equations (solid lines) for the linear core system, at
various densities and $\chi=1$. The low-density approximation corresponding
to Eq. (\ref{SHSD_g2_low}) is also shown. (\emph{Right}) The corresponding
static structure factors from SHSD simulations (solid symbols, average
of ten snapshots of a system with $N=10^{5}$ particles in a cubic
periodic box) and HNC calculations (solid lines). The time step was
kept sufficiently small in the SHSD simulations to ensure that the
results are faithfully represent the SHSD fluid with time-step errors
smaller than the statistical uncertainty.}

\end{figure}

Remarkably, we find \emph{numerically} that the effective potential
(\ref{SHSD_U_eff}) can predict exactly $g_{2}(x)$ at \emph{all}
densities. In fact, we have numerically observed that the SHSD fluid
behaves thermodynamically \emph{identically} to a system of penetrable
spheres interacting with a linear core pairwise potential for all
$\phi$ and $\chi$. Figure \ref{SHSD_g2} shows a comparison between
the pair correlation function of the SHSD fluid on one hand, and a
Monte Carlo calculation using the linear core pair potential on the
other, at several densities. Also shown is a numerical solution to
the hypernetted chain (HNC) integral equations for the linear core
system, inspired by its success for the Gaussian core model \citet{GaussianCoreHNC}.
The excellent agreement at all densities permits the use of the HNC
result in practical applications, notably the calculation of the transport
coefficients via the Enskog-like kinetic theory presented in Section
\ref{SectionEnskogEq}. We also show the static structure factor $S(k)$
in Fig. \ref{SHSD_g2}, and find very good agreement between numerical
results and the HNC theory, as expected since $S(k)$ can be expressed
as the Fourier transform of $h(r)=g_{2}(r)-1$.

For collision frequencies $\chi\lesssim1$ the structure of the SHSD
fluid is quite different from that of the hard-sphere fluid because
the particles inter-penetrate and overlap significantly. Interestingly,
in the limit of infinite collision frequency $\chi\rightarrow\infty$
the SHSD fluid reduces to the hard-sphere (HS) fluid for sufficiently
low densities. In fact, if the density $\phi$ is smaller than the
freezing point for the HS system, the structure of the SHSD fluid
approaches, as $\chi$ increases, that of the HS fluid. For higher
densities, if $\chi$ is sufficiently high, crystallization is observed
in SHSD, either to the usual hard-sphere crystals if $\phi$ is lower
than the close-packing density, or if not, to an unusual partially
ordered state with multiple occupancy per site, typical of weakly
repulsive potentials \citet{MultipleOccupancyCrystal}. Monte Carlo
simulations of the linear core penetrable sphere system show identical
freezing behavior with SHSD, confirming the surprising equivalence
even at non-fluid densities. This points to a conjecture that the
(unique) stationary solution to the BBGKY hierarchy (\ref{BBGKY_df_dt})
is the equilibrium Gibbs distribution,\[
f_{s}^{E}=\frac{\prod_{i=1}^{s}M(\V{v}_{i})}{Z_{N}}\int_{\V{r}_{s+1}}\ldots\int_{\V{r}_{N}}\exp\left[-\beta\sum_{i<j}U_{eff}(r_{ij})\right]d\V{r}_{s+1}\ldots d\V{r}_{N},\]
where $M$ is a Maxwellian.

\section{\label{sec:Results}Results}

In this Section we perform several numerical experiments with the
SHSD algorithm. Firstly, we compare the theoretical predictions for
the transport properties of the SHSD fluid based on the HNC theory
for the linear core penetrable sphere system with results from particle
simulations. We then compute dynamic structure factors and compare
them to predictions of fluctuating hydrodynamics. Finally, we study
the motion of a Brownian bead suspended in an SHSD fluid.

\subsection{Transport Coefficients}

\begin{figure}[tbph]
\begin{centering}
\includegraphics[width=0.495\columnwidth]{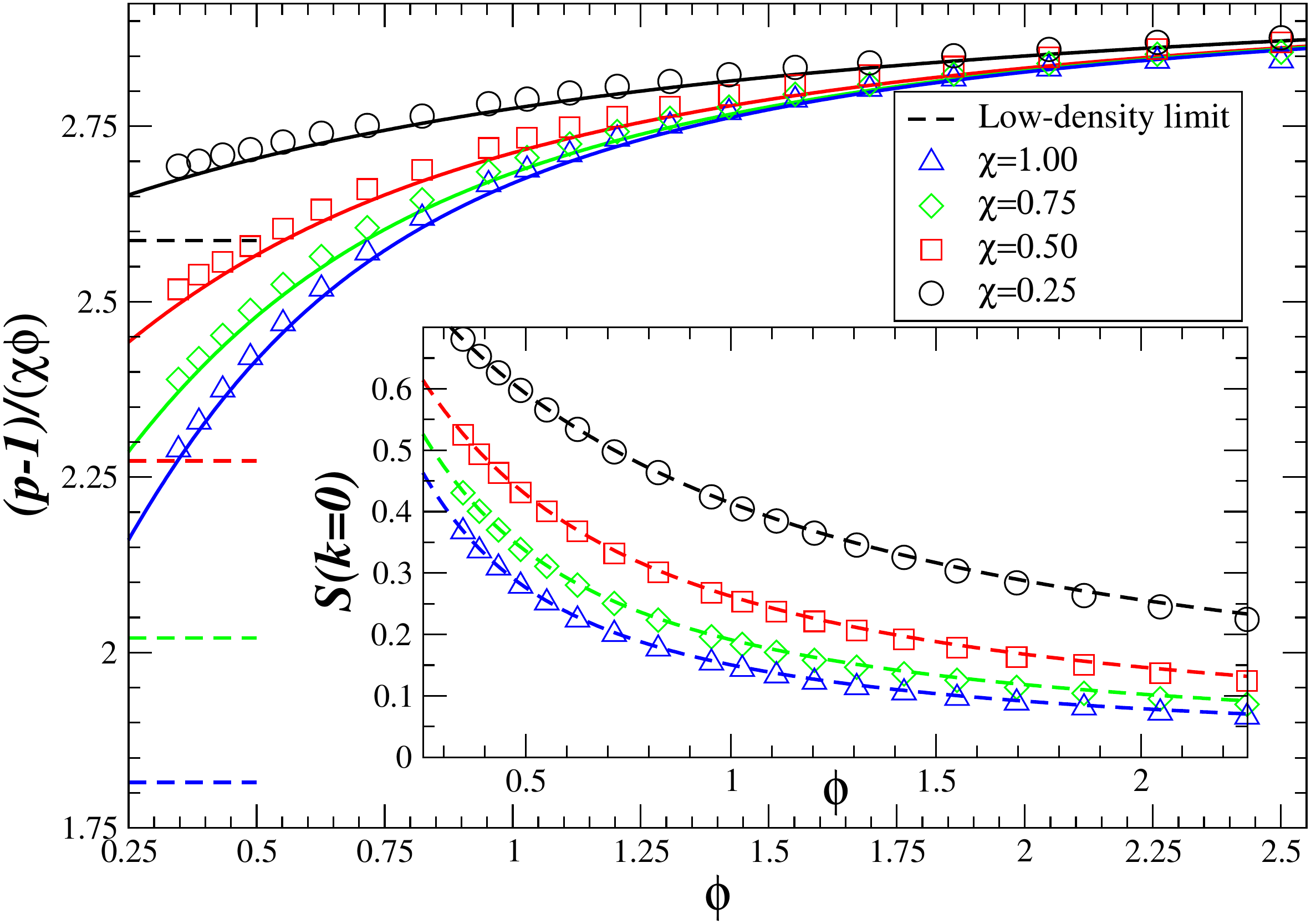}\includegraphics[width=0.495\columnwidth]{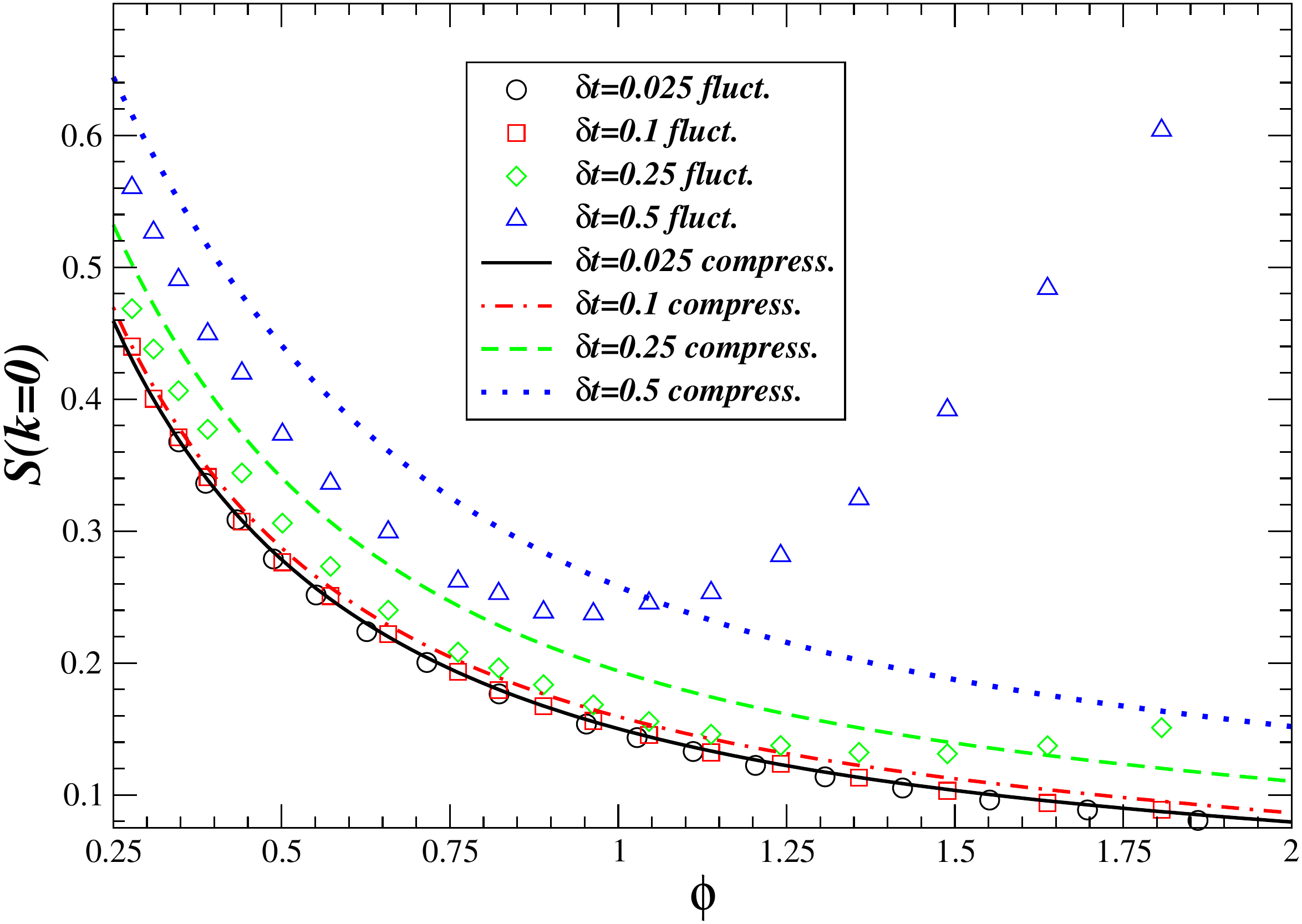}
\par\end{centering}

\caption{\label{SHSD_EOS_consistency}(\emph{Left}) Normalized equation of
state $(p-1)/(\chi\phi)$ for the SHSD fluid at several cross-section
factors $\chi$ (different symbols, $N=10^{5}$ particles in a cubic
box) compared to theoretical predictions based on the virial theorem
(\ref{SHSD_p}) with the HNC approximation to $g_{2}(x)$ (solid lines).
The inset compares the compressibility $(p+\phi dp/d\phi)^{-1}$ (dashed
lines) to the structure factor at the origin $S(k\rightarrow0)$ (symbols),
measured using a direct Fourier transform of the particle positions
for small $k$ and extrapolating to $k=0$. The dimensionless time
step $\delta t=0.025$ is kept constant and small as the density is
changed. (\emph{Right}) Thermodynamic consistency between the compressibility
(lines) and the large-scale density fluctuations $S(k\rightarrow0)$
(symbols) for different dimensionless time steps $\delta t$, keeping
$\chi=1$ fixed.}

\end{figure}

The equation of state of the SHSD fluid for a given $\chi$ is $P=p(\phi)Nk_{B}T/V$,
where $p(\phi)$ is given in Eq. (\ref{SHSD_p}). According to statistical
mechanics, the structure factor at the origin is equal to the isothermal
compressibility, that is, \[
S_{0}=S(\omega=0,k=0)=\tilde{c}_{T}^{-2}=(p+\phi dp/d\phi)^{-1}\]
where $c_{T}=\tilde{c}_{T}\sqrt{k_{B}T/m}$ is the isothermal speed
of sound. In the inset in the top part of Fig. \ref{SHSD_EOS_consistency},
we directly demonstrate the thermodynamic consistency of SHSD by comparing
the compressibility calculated through numerical differentiation of
the pressure, to the structure factor at the origin. The pressure
is easily measured in the particle simulations by keeping track of
the total collisional momentum exchange during a long period, and
its derivative was obtained by numerical differentiation. The structure
factor is obtained through a temporal average of a Fast Fourier Transform
approximation to the discrete Fourier Transform of the particle positions
$\left\Vert \sum_{i}\exp(-i\V{k}\cdot\V{r}_{i})\right\Vert ^{2}$.
The value $S(k=0)$ is estimated by fitting a parabolic dependence
for small $k$ and extrapolating to $k=0$.

As pointed out earlier, the dimensionless time step $\delta t=D/\sqrt{k_{B}T_{0}/m}$
in the SHSD algorithm should be kept reasonably small, $\delta t\ll\min\left[1,(\phi\chi)^{-1}\right]$,
in order to faithfully simulate the SHSD fluid. As the time step becomes
too large we expect to see deviations from the correspondence with
the linear core system and thus a violation of thermodynamic consistency.
This is indeed observed in our numerical results, shown in Fig. \ref{SHSD_EOS_consistency},
where we compare the structure factor at the origin as estimated through
the equation of state with that obtained from a direct Fourier transform
of the particle positions. We should point out that when discussing
thermodynamic consistency one has to define what is meant by the the
derivative $dp/d\phi$. We choose to keep the collisional frequency
prefactor $\chi$ and the \emph{dimensionless} time step $\delta t$
constant as we change the density, that is, we study the thermodynamic
consistency of a \emph{time-discrete SHSD fluid} defined by the parameters
$\phi$, $\chi$ \emph{and} $\delta t$. The results in Fig. \ref{SHSD_EOS_consistency}
show that there are significant deviations from thermodynamic consistency
when the average number of collisions per particle per time step is
larger than one. This happens at the highest densities for $\delta t\gtrsim0.1$,
but is not a problem at the lowest densities. Nevertheless, a visible
inconsistency is observed even at the lower densities for $\delta t\gtrsim0.25$,
which comes because particles travel too far compared to their own
size during a time step.

\begin{figure}[tbph]
\begin{centering}
\includegraphics[width=0.495\textwidth]{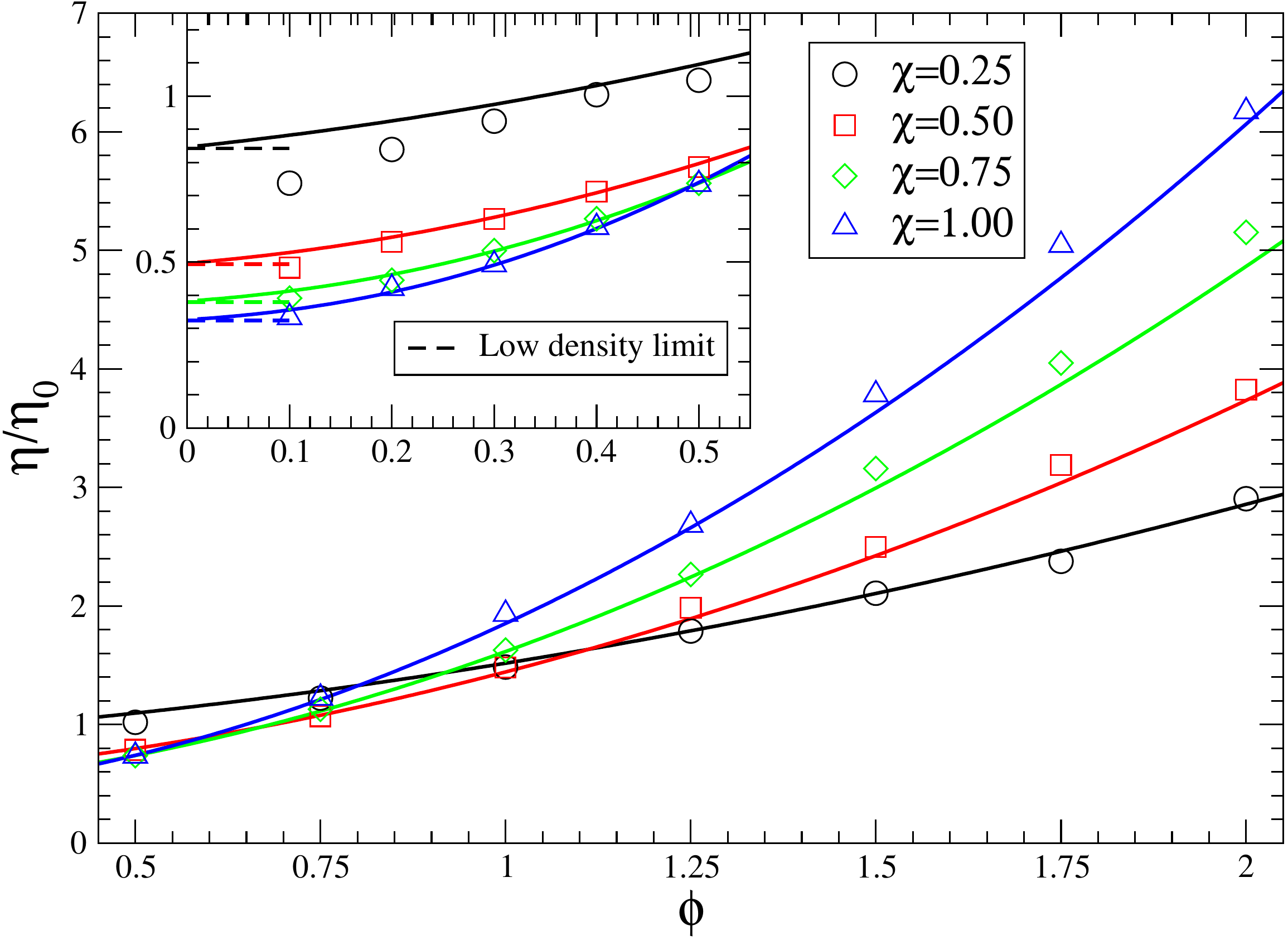}\includegraphics[width=0.495\textwidth]{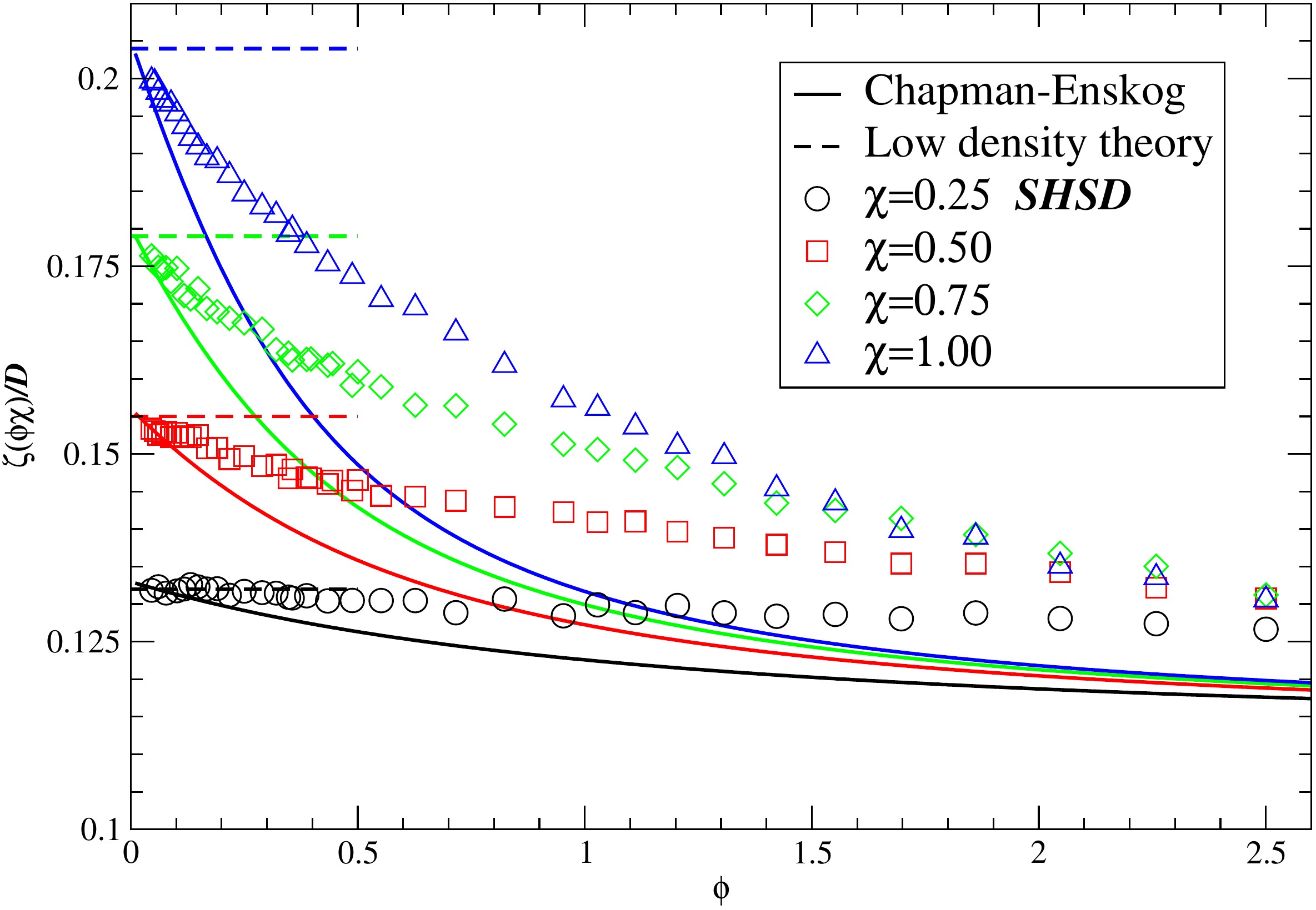}
\par\end{centering}

\caption{\label{SHSD_transport_coeffs}Comparison between numerical results
for SHSD at several collision frequencies (different symbols) with
predictions based on the stochastic Enskog equation using the HNC
approximation for $g_{2}(x)$ (solid lines). The low-density approximations
are also indicated (dashed lines). (\emph{Left}) The normalized shear
viscosity $\eta/\eta_{0}$ at high and low densities (inset), as measured
using an externally-forced Poiseuille flow. There are significant
corrections (Knudsen regime) for large mean free paths (i.e., at low
densities and low collision rates). (\emph{Right}) The normalized
diffusion coefficient $\zeta(\chi\phi)/\zeta_{0}$, as measured from
the mean square displacement of the particles. The time step was kept
sufficiently small in the SHSD simulations to ensure that the results
are faithfully represent the SHSD fluid with time-step errors smaller
than the statistical and measurement errors.}

\end{figure}

Having established that the HNC closure provides an excellent approximation
$g_{2}^{(HNC)}\approx g_{2}$ for the pair correlation function of
the SHSD fluid, we can obtain estimates for the transport coefficients
by calculating the first four moments of $g_{2}^{(HNC)}(x)$ and substituting
them in the results of the Enskog kinetic theory presented in Section
\ref{SectionEnskogEq}. In Figure \ref{SHSD_transport_coeffs} we
compare the theoretical predictions for the diffusion coefficient
$\zeta$ and the viscosity $\eta$ to the ones directly calculated
from SHSD particle simulations. We measure $\zeta$ directly from
the average mean square displacement of the particles. We estimate
$\eta$ by calculating the mean flow rate in Poiseuille parabolic
flow between two thermal hard walls due to an applied constant force
on the particles%
\footnote{Similar results are obtained by calculating the viscous contributions
to the kinetic and collisional stress tensor in non-equilibrium simulations
of Couette shear flow. This kind of calculation additionally gives
the split in the viscosity between kinetic and collisional contributions.%
}. Surprisingly, good agreement is found for the shear viscosity at
all densities. Similar matching was observed for the thermal conductivity
$\kappa$. The corresponding results for the diffusion coefficient
show significant ($\sim25\%$) deviations for the self-diffusion coefficient
at higher densities because of larger corrections due to higher-order
correlations.

\subsection{Dynamic Structure Factors}

The hydrodynamics of the spontaneous thermal fluctuations in the SHSD
fluid is expected to be described by the Landau-Lifshitz Navier-Stokes
(LLNS) equations for the fluctuating field $\V{U}=(\rho_{0}+\delta\rho,\delta v,T_{0}+\delta T)$
linearized around a reference equilibrium state $\V{U}_{0}=(\rho_{0},\V{v}_{0}=\V{0},T_{0})$
\citet{Landau:Fluid,FluctHydroNonEq_Book}. For the SHSD fluid the
linearized equation of state is \[
P=p(\phi)\frac{Nk_{B}T}{V}\approx(p_{0}+\tilde{c}_{T}^{2}\frac{\delta\rho}{\rho_{0}}+p_{0}\frac{\delta T}{T_{0}})\rho_{0}c_{0}^{2},\]
and there is no internal energy contribution to the energy density,\[
e\approx\frac{3}{2}\frac{Nk_{B}T}{V}=e_{0}+c_{v}T_{0}\delta\rho+\rho_{0}c_{v}\delta T,\]
where $p_{0}=p(\phi_{0})$, $c_{0}=k_{B}T/m$, and $c_{v}=3k_{B}/2m$,
giving an adiabatic speed of sound $c_{s}=\tilde{c}_{s}c_{0},$ where
$\tilde{c}_{s}^{2}=\tilde{c}_{T}^{2}+2p^{2}/3$. Omitting the $\delta$'s
for notational simplicity, for one-dimensional flows the LLNS equations
take the form

\begin{equation}
\left[\begin{array}{c}
\partial_{t}\rho\\
\partial_{t}v\\
\partial_{t}T\end{array}\right]=-\frac{\partial}{\partial x}\left[\begin{array}{c}
\rho_{0}v\\
c_{T}^{2}\rho_{0}^{-1}\rho+p_{0}c_{0}^{2}T_{0}^{-1}T\\
p_{0}c_{0}^{2}c_{v}^{-1}v\end{array}\right]+\frac{\partial}{\partial x}\left[\begin{array}{c}
0\\
\rho_{0}^{-1}\eta_{0}v_{x}\\
\rho_{0}^{-1}c_{v}^{-1}\kappa_{0}T_{x}\end{array}\right]+\frac{\partial}{\partial x}\left[\begin{array}{c}
0\\
\rho_{0}^{-1}\sqrt{2\eta_{0}k_{B}T_{0}}W^{(v)}\\
\rho_{0}^{-1}c_{v}^{-1}T_{0}\sqrt{2\kappa_{0}k_{B}}W^{(T)}\end{array}\right],\label{LLNS_1D_SHSD}\end{equation}
where $W^{(v)}$ and $W^{(T)}$ are independent spatio-temporal white
noise Gaussian fields.

By solving these equations in the Fourier wavevector-frequency domain
for $\widehat{\V{U}}(k,\omega)$ and performing an ensemble average
over the fluctuating stresses we can obtain the equilibrium (stationary)
spatio-temporal correlations (covariance) of the fluctuating fields.
We express these correlations in terms of the $3\times3$ symmetric
positive-definite \emph{hydrodynamic structure factor matrix} $\M{S}_{H}(k,\omega)=\left\langle \widehat{\V{U}}\widehat{\V{U}}^{\star}\right\rangle $
\citet{LLNS_S_k}. The \emph{hydrostatic structure factor matrix}
$\M{S}_{H}(k)$ is obtained by integrating $\M{S}_{H}(k,\omega)$
over all frequencies, \begin{equation}
\M{S}_{H}(k)=\left[\begin{array}{ccc}
\rho_{0}c_{T}^{-2}k_{B}T_{0} & 0 & 0\\
0 & \rho_{0}^{-1}k_{B}T_{0} & 0\\
0 & 0 & \rho_{0}^{-1}c_{v}^{-1}k_{B}T_{0}^{2}\end{array}\right].\label{S_U_continuum}\end{equation}
We use $\M{S}_{H}(k)$ for an ideal gas (i.e., for $p_{0}=1$, $\tilde{c}_{T}=1$)
to non-dimensionalize $\M{S}_{H}(k,\omega)$, for example, we express
the spatio-temporal cross-correlation between density and velocity
through the dimensionless hydrodynamic structure factor\[
S_{\rho,v}(k,\omega)=\left(\rho_{0}c_{0}^{-2}k_{B}T_{0}\right)^{-\frac{1}{2}}\left(\rho_{0}^{-1}k_{B}T_{0}\right)^{-\frac{1}{2}}\left\langle \hat{\rho}(k,\omega)\hat{v}^{\star}(k,\omega)\right\rangle .\]
For the non-ideal SHSD fluid the density fluctuations have a spectrum
\[
S_{\rho}(k)=\left(\rho_{0}c_{0}^{-2}k_{B}T_{0}\right)^{-1}\left\langle \hat{\rho}(k)\hat{\rho}^{\star}(k)\right\rangle =\tilde{c}_{T}^{-2},\]
which only captures the small $k$ behavior of the full (particle)
structure factor $S(k)$ (see Fig. \ref{SHSD_g2}), as expected of
a continuum theory that does not account for the structure of the
fluid. Typically only the density-density dynamic structure factor
is considered because it is accessible experimentally via light scattering
measurements and thus most familiar. However, in order to fully access
the validity of the full LLNS system one should examine the dynamic
correlations among all pairs of variables. The off-diagonal elements
of the static structure factor matrix $\M{S}_{H}(k)$ vanish because
the primitive hydrodynamic variables are instantaneously uncorrelated,
however, they have non-trivial dynamic correlations visible in the
off-diagonal elements of the dynamic structure factor matrix $\M{S}_{H}(k,\omega)$.

\begin{figure}[tbph]
\begin{centering}
\includegraphics[width=0.7\textwidth]{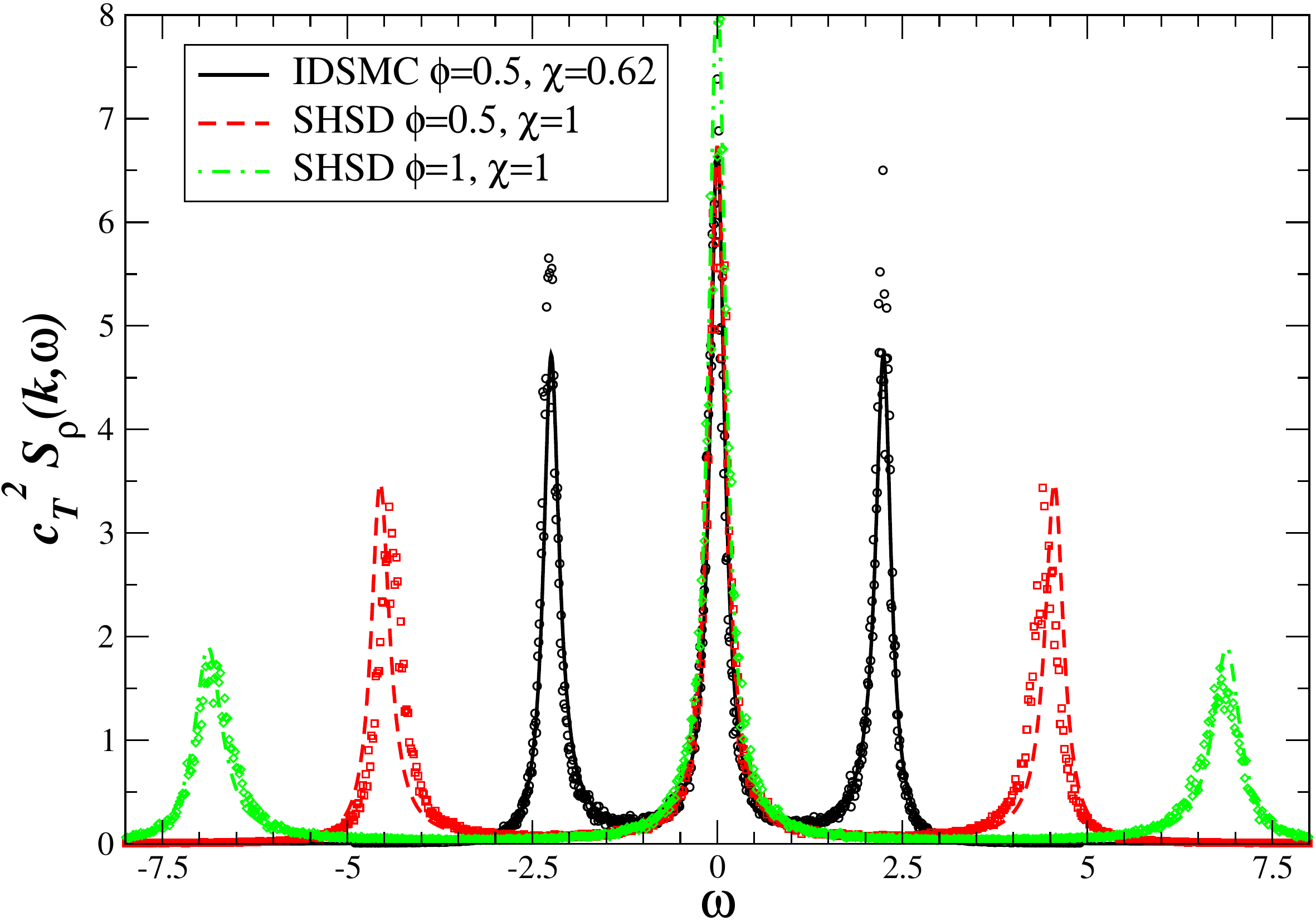}
\par\end{centering}

\caption{\label{S_kw_SHSD}Normalized density fluctuations $\tilde{c}_{T}^{2}S_{\rho}(k,\omega)$
for $kD\approx0.070$ for an ideal Maxwell I-DSMC ($\phi=0.5$, $\chi=0.62$)
and two non-ideal SHSD ($\phi=0.5$, $\chi=1$ and $\phi=1$, $\chi=1$)
fluids of similar kinematic viscosity, as obtained from particle simulations
(symbols with parameters , $k_{B}T_{0}=1$, $m=1$). The predictions
of the LLNS equations are also shown for comparison in the same color
(solid lines). For the SHSD fluid we obtained the transport coefficients
from the Enskog theory with the HNC approximation to $g_{2}$, while
for the Maxwell I-DSMC fluid we numerically estimated the viscosity
and thermal conductivity. }

\end{figure}

\begin{figure}[tbph]
\begin{centering}
\includegraphics[width=0.45\textwidth]{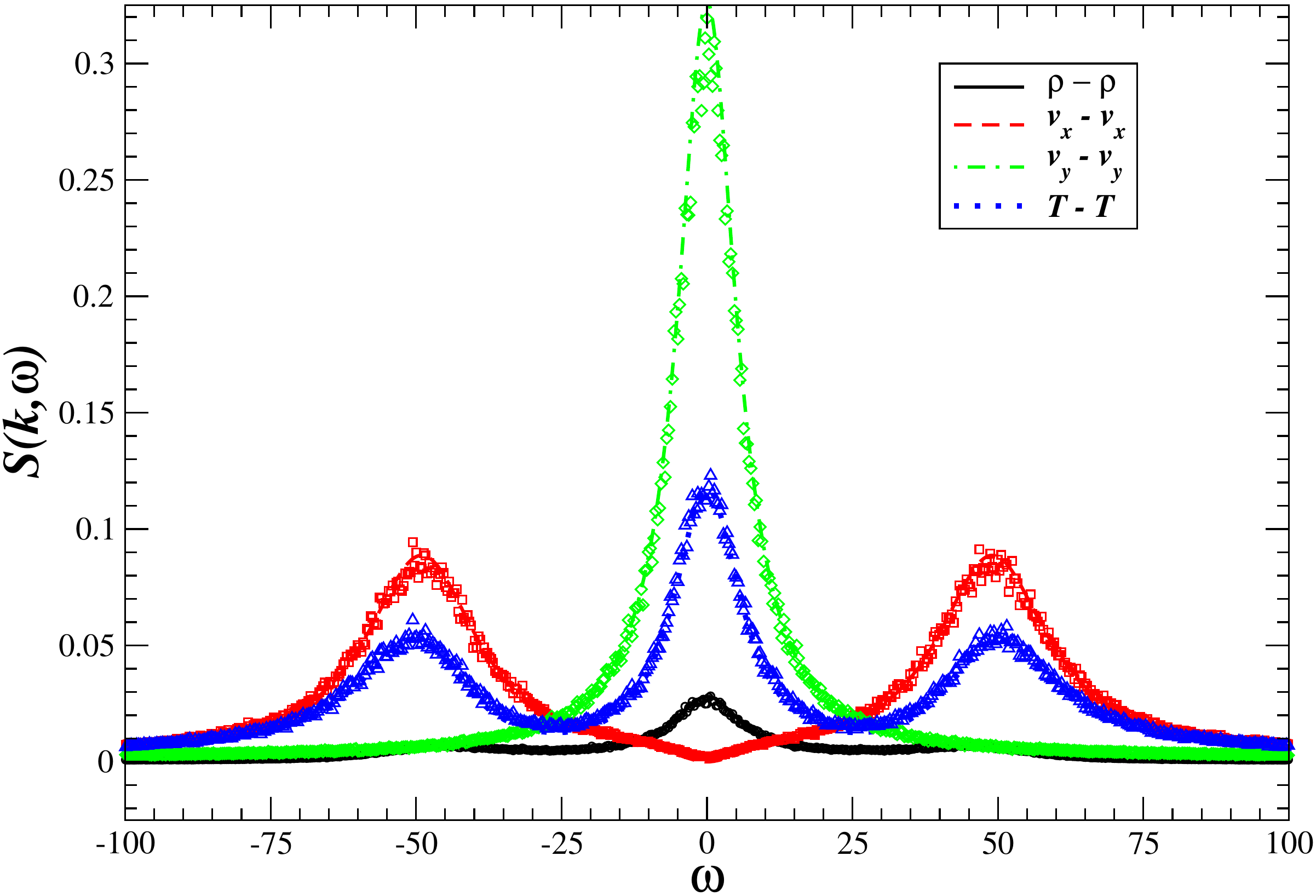}\includegraphics[width=0.45\textwidth]{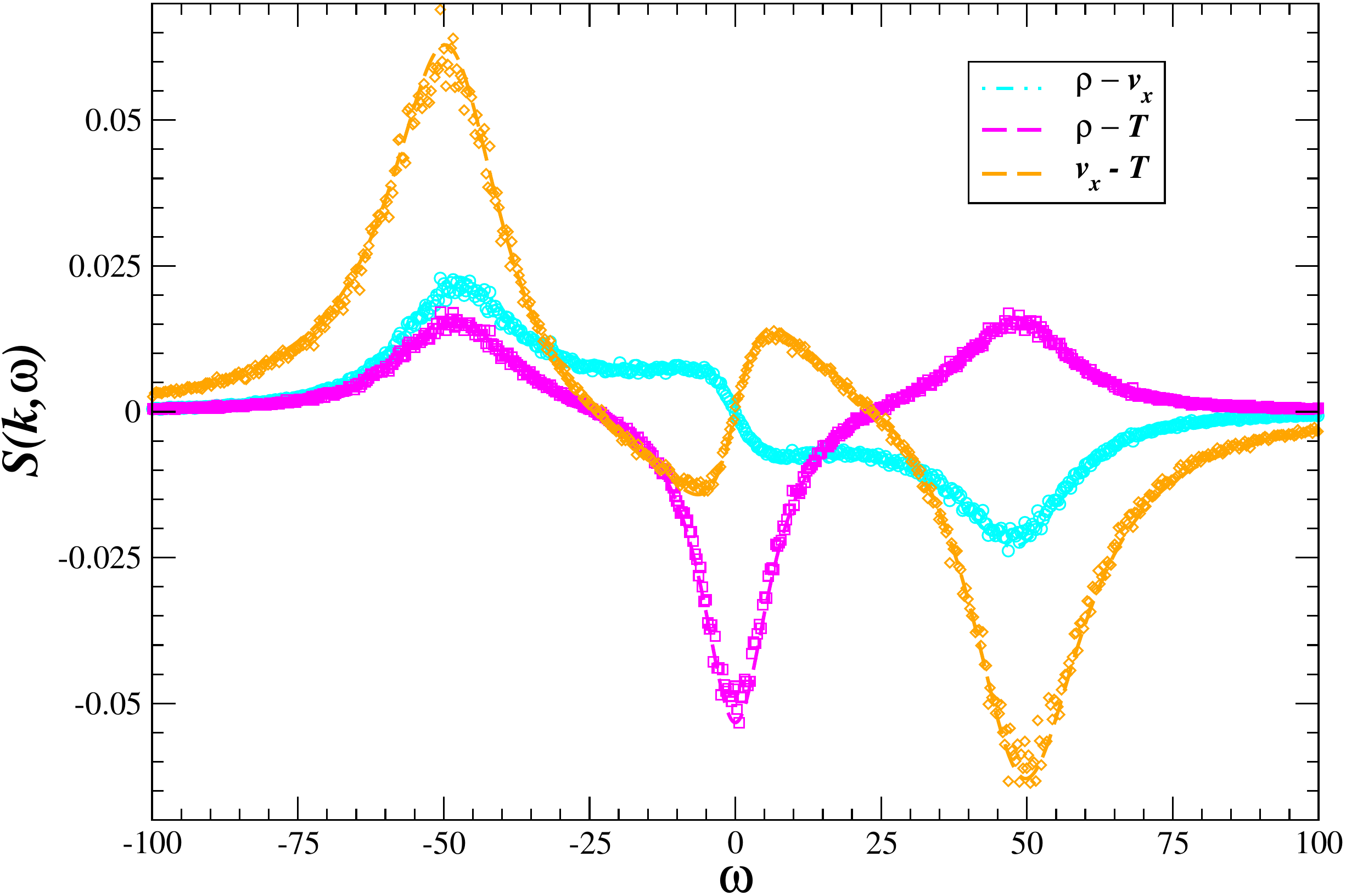}
\par\end{centering}

\caption{\label{S_kw_matrix_SHSD}Selected diagonal (left panel) and off-diagonal
elements (right panel) of the non-dimensionalized hydrodynamic structure
factor matrix $\M{S}_{H}(k,\omega)$ for a large wavenumber $kD\approx0.50$
for an SHSD fluid at $\phi=1$, $\chi=1$ (symbols), compared to the
predictions from the LLNS equations (lines of same color). The remaining
parameters are as in Fig. \ref{S_kw_SHSD}.}

\end{figure}

In Figs. \ref{S_kw_SHSD} and \ref{S_kw_matrix_SHSD} we compare theoretical
and numerical results the hydrodynamic structure factors for the SHSD
fluid with $\chi=1$ at two densities for a small and a medium $k$
value {[}$kD/(2\pi)\approx0.01\mbox{ and }0.08$]. In this figure
we show selected elements of $\M{S}_{H}(k,\omega)$ as predicted by
the analytical solution to Eqs. (\ref{LLNS_1D_SHSD}) with parameters
obtained by using the HNC approximation to $g_{2}$ in the Enskog
kinetic theory presented in Section \ref{SectionEnskogEq}. Therefore,
for SHSD the theoretical calculations of $\M{S}_{H}(k,\omega)$ do
not use any numerical inputs from the particle runs. We also show
hydrodynamic structure factors obtained from particle simulations
in a quasi-one-dimensional setup in which the simulation cell was
periodic and long along the $x$ axis, and divided into 60 hydrodynamic
cells of length $5D$. Finite-volume averages of the hydrodynamic
conserved variables were then calculated for each cell every $10$
time steps and a Fast Fourier Transform used to obtain hydrodynamic
structure factors for several wavenumbers. Figure \ref{S_kw_SHSD}
shows very good agreement between theory and numerics, and clearly
shows the shifting of the two symmetric Brillouin peaks at $\omega\approx c_{s}k$
toward higher frequencies as the compressibility of the SHSD fluid
is reduced and the speed of sound increased. Figure \ref{S_kw_matrix_SHSD}
shows that the positions and widths of the side Brillouin peaks and
the width of the central Rayleigh are well-predicted for all elements
of $\M{S}_{H}(k,\omega)$ for a wide range of $k$ values, demonstrating
that the SHSD fluid shows the expected fluctuating hydrodynamic behavior.

\subsection{Brownian Walker VACF}

As an illustration of the correct hydrodynamic behavior of the SHSD
fluid and the significance of compressibility, we study the velocity
autocorrelation function (VACF) $C(t)=\left\langle v_{x}(0)v_{x}(t)\right\rangle $
for a single neutrally-buoyant hard sphere Brownian bead of mass $M$
and radius $R$ suspended in an SHSD fluid of mass density $\rho$.
This problem is relevant to the modeling of polymer chains or (nano)colloids
in solution, and led to the discovery of a long power-law tail in
$C(t)$ \citet{VACF_Alder} which has since become a standard test
for hydrodynamic behavior of solvents \citet{BrownianLB_VACF,MPCD_VACF,BrownianSRD_Review}.
Here the fluid particles interact via stochastic collisions, exactly
as in I-DSMC. The interaction between fluid particles and the bead
is treated as if the SHSD particles are hard spheres of diameter $D_{s}$,
chosen to be somewhat smaller than their interaction diameter with
other fluid particles (specifically, we use $D_{s}=D/4$) for computational
efficiency reasons, using an event-driven algorithm \citet{DSMC_AED}.
Upon collision with the bead the relative velocity of the fluid particle
is reversed in order to provide a no-slip condition at the surface
of the suspended sphere \citet{BrownianSRD_Review,DSMC_AED} (slip
boundaries give qualitatively identical results). For comparison,
an ideal I-DSMC fluid of comparable viscosity is also simulated.

Theoretically, $C(t)$ has been calculated from the linearized (compressible)
fluctuating Navier-Stokes (NS) equations \citet{BrownianSRD_Review}.
The results are analytically complex even in the Laplace domain, however,
at short times an inviscid compressible approximation applies. At
large times the compressibility does not play a role and the incompressible
NS equations can be used to predict the long-time tail. At short times,
$t<t_{c}=2R/c_{s}$, the major effect of compressibility is that sound
waves generated by the motion of the suspended particle carry away
a fraction of the momentum, so that the VACF quickly decays from its
initial value $C(0)=k_{B}T/M$ to $C(t_{c})\approx k_{B}T/M_{eff}$,
where $M_{eff}=M+2\pi R^{3}\rho/3$. At long times, $t>t_{visc}=4\rho R_{H}^{2}/3\eta$,
the VACF decays as in an incompressible fluid, with an asymptotic
power-law tail $(k_{B}T/M)(8\sqrt{3\pi})^{-1}(t/t_{visc})^{-3/2}$,
in disagreement with predictions based on the Langevin equation (Brownian
dynamics), $C(t)=(k_{B}T/M)\exp\left(-6\pi R_{H}\eta t/M\right)$.
We have estimated the effective (hydrodynamic) colloid radius $R_{H}$
from numerical measurements of the Stokes friction force $F=-6\pi R_{H}\eta v$
and found it to be somewhat larger than the hard-core collision radius
$R+D_{s}/2$, but for the calculations below we use $R_{H}=R+D_{s}/2$.

\begin{figure}[tbph]
\begin{centering}
\includegraphics[width=0.75\textwidth]{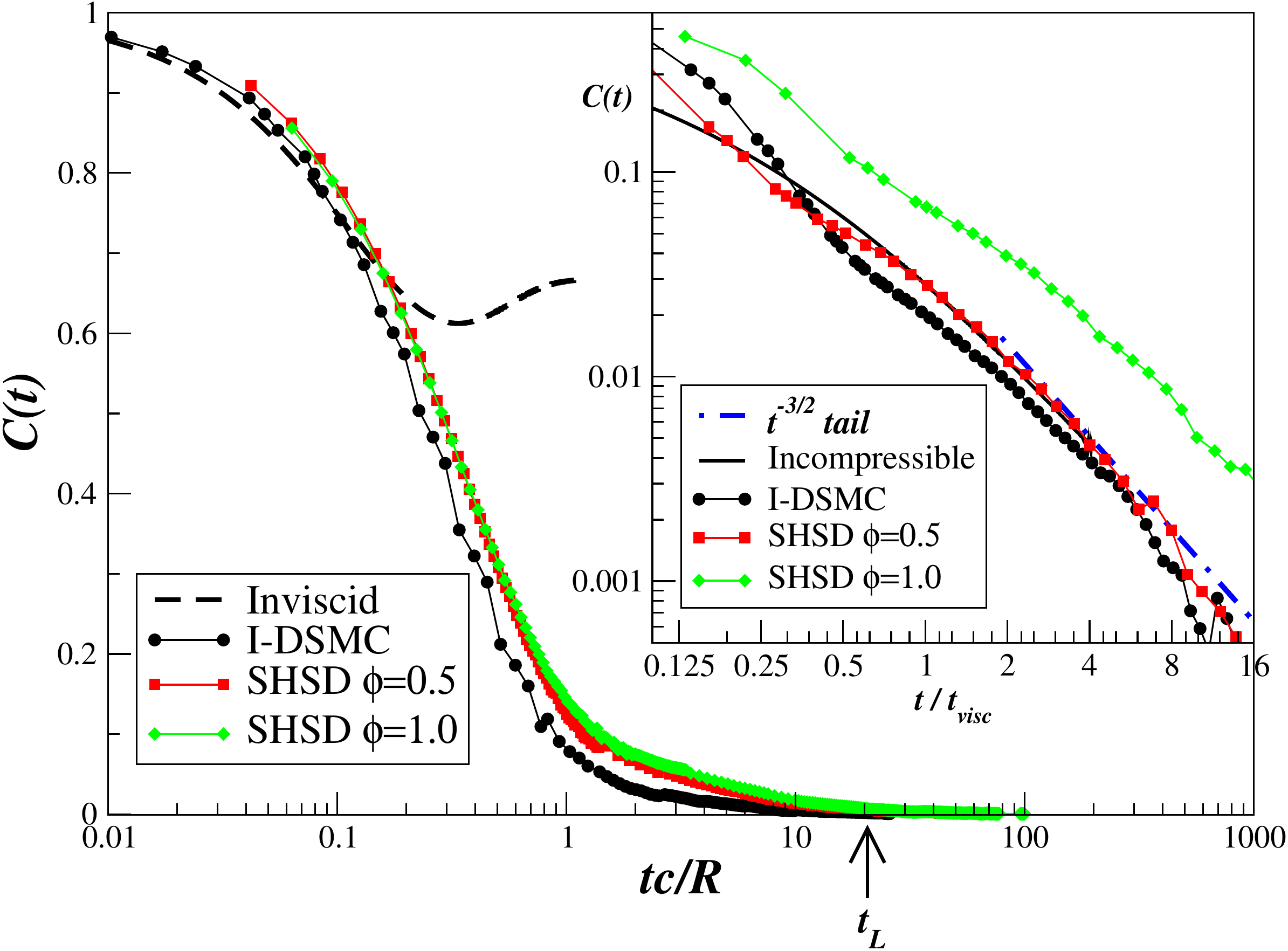}
\par\end{centering}

\caption{\label{VACF_SHSD}The velocity autocorrelation function for a neutrally
buoyant hard sphere suspended in a non-ideal SHSD ($\chi=1$) fluid
at two densities (symbols), $\phi=0.5$ and $\phi=1.0$, as well as
an ideal Maxwell I-DSMC fluid ($\phi=0.5$, $\chi=0.62$, symbols),
at short and long times (inset). For the more compressible (less viscous)
fluids the long time tails are statistically measurable only up to
$t/t_{visc}\approx5$. The theoretical predictions based on the inviscid,
for short times, or incompressible, for long times, Navier-Stokes
equations are also shown (lines). }

\end{figure}

In Fig. \ref{VACF_SHSD} numerical results for the VACF in a Maxwell
I-DSMC fluid and an SHSD fluid at two different densities are compared
to the theoretical predictions. The diameter of the nano-colloidal
particle is only $2.5D$ (i.e., $R_{H}=1.375D$), although we have
performed simulations using larger spheres as well with very similar
(but less accurate) results. Since periodic boundary conditions were
used we only show the tail up to about the time at which sound waves
generated by its periodic images reach the particle, $t_{L}=L/c_{s}$,
where the simulation box was $L=25D$. In dimensionless units, the
viscosity $\eta=\tilde{\eta}D^{-2}\sqrt{mk_{B}T}$ was measured to
be $\tilde{\eta}\approx0.75$ for both the Maxwell I-DSMC fluids and
the SHSD fluid at $\phi=0.5$, and $\tilde{\eta}\approx1.9$ for SHSD
at $\phi=1$. The results in Fig. \ref{VACF_SHSD} are averages over
10 runs, each of length $T/t_{visc}\approx2\cdot10^{5}$ for I-DSMC,
$T/t_{visc}\approx1\cdot10^{5}$ for SHSD at $\phi=0.5$, and $T/t_{visc}\approx4.5\cdot10^{4}$
for SHSD at $\phi=1.0$, where in atomistic time units $t_{0}=D\sqrt{m/k_{B}T}$
the viscous time scale is $t_{visc}/t_{0}\approx6\phi/(3\pi\tilde{\eta}).$

It is seen, as predicted, that the compressibility or the sound speed
$c_{s}$, determines the early decay of the VACF. The exponent of
the power-law decay at large times is also in agreement with the hydrodynamic
predictions. The coefficient of the VACF tail agrees reasonably well
with the hydrodynamic prediction for the less dense fluids, however,
there is a significant deviation of the coefficient for the densest
fluids, perhaps due to ordering of the fluid around the suspended
sphere, not accounted for in continuum theory. In order to study this
discrepancy in further detail one would need to perform simulations
with a much larger bead. This is prohibitively expensive with the
serial event-driven algorithm used here \citet{DSMC_AED} and requires
either parallelizing the code or using a hybrid particle-continuum
method \citet{DSMC_Hybrid}, which we leave for future work.

\section{Conclusions}

We have successfully generalized the traditional DSMC algorithm for
simulating rare gas flows to flows of dense non-ideal fluids. Constructing
such a thermodynamically-consistent Stochastic Hard Sphere Dynamics
(SHSD) algorithm required first eliminating the grid artifacts from
traditional DSMC. These artifacts are small in traditional DSMC simulations
of rarefied gases because the collisional cell size is kept significantly
smaller than the mean free path \citet{DSMC_CellSizeError}, but become
pronounced when dense flows are simulated because the collisional-stress
tensor is not isotropic. Our Isotropic DSMC (I-DSMC) method is a grid
free DSMC variant with pairwise spherically-symmetric \emph{stochastic}
interactions between the particles, just as classical fluids simulated
by molecular dynamics (MD) use a pairwise spherically-symmetric \emph{deterministic}
interaction potential. The I-DSMC method can therefore be viewed as
a transition from the DSMC method, suitable for rarefied flows, to
the MD method, suitable for simulating dense liquids (and solids).

It has long been apparent that manipulating the stochastic collision
rules in DSMC can lead to a wide range of fluid models, including
non-ideal ones \citet{DSMC_DenseFluids,DSMC_CBA}. It has also been
realized that DSMC, as a kinetic Monte Carlo method, is not limited
to solving the Boltzmann equation \citet{DSMC_Granular2} but can
be generalized to Enskog-like kinetic equations \citet{DSMC_Enskog_Frezzotti,DSMC_Enskog}.
However, what has been so far elusive is to construct a DSMC collision
model that is thermodynamically consistent, meaning that the resulting
fluid structure and the equation of state are consistent with each
other as required by statistical mechanics. We overcame this hurdle
here by constructing stochastic collision kernels in I-DSMC to be
as close as possible to those of the classical hard-sphere deterministic
system. Thus, in the SHSD algorithm randomly chosen pairs of approaching
and overlapping particles undergo collisions as if they were hard
spheres of variable diameter. This is similar to the modified collision
rules used to construct a consistent non-ideal Multi Particle Collision
Dynamics fluid in Refs. \citet{MPCD_CBA,MPCD_CBA_2}.

We demonstrated the consistent thermodynamic behavior of the SHSD
system by observing that it has identical structure and thermodynamic
properties to a Hamiltonian system of penetrable spheres interacting
with a linear core potential, even up to solid densities. We found
that at fluid densities the pair correlation function $g_{2}(r)$
of the linear core system is well-described by the approximate HNC
closure, enabling us to obtain moments of $g_{2}(r)$. These moments
were then used as inputs in a modified Chapman-Enskog calculation
to obtain excellent estimates of the equation of state and transport
coefficients of the SHSD fluid over a wide range of densities. We
do not yet have a complete theoretical understanding of our surprising
finding that the SHSD system behaves thermodynamically identically
to the linear core system. An important open question remains whether
by choosing a different collision kernel one can obtain stochastic
fluids corresponding to Hamiltonian systems of penetrable spheres
interacting with effective pair potentials $U_{eff}(r)$ other than
the linear core potential.

The SHSD algorithm is similar in nature to DPD and has a similar computational
complexity. The essential difference is that DPD has a continuous-time
formulation (a system of stochastic ODEs), where as the SHSD dynamics
is discontinuous in time (Master Equation). This is similar to the
difference between MD for continuous potentials and discontinuous
potentials. Just as DSMC is a stochastic alternative to hard-sphere
MD for low-density gases, SHSD is a stochastic modification of hard-sphere
MD for dense gases. On the other hand, DPD is a modification of MD
for smooth potentials to allow for larger time-steps and a conservative
thermostat.

A limitation of SHSD is that for reasonable values of the collision
frequency ($\chi\sim1$) and density ($\phi\sim1$) the fluid is still
relatively compressible compared to a dense liquid, $S(k=0)=\tilde{c}_{T}^{-2}>0.1$.
Indicative of this is that the diffusion coefficient is large relative
to the viscosity as it is in typical DPD simulations, so that the
Schmidt number $S_{c}=\eta(\rho\zeta)^{-1}$ is less than 10 instead
of being on the order of 100-1000. Achieving higher $\tilde{c}_{T}$
or $S_{c}$ requires high collision rates (for example, $\chi\sim10^{4}$
is used in Ref. \citet{DPDSchmidtNumbers}) and appropriately smaller
time steps to ensure that there is at most one collision per particle
per time step, and this requires a similar computational effort as
in hard-sphere molecular dynamics at a comparable density. At low
and moderate gas densities the SHSD algorithm is not as efficient
as DSMC at a comparable collision rate. However, for a wide range
of compressibilities, SHSD is several times faster than the alternative
deterministic Event-Driven MD (EDMD) for hard spheres \citet{EventDriven_Alder,AED_Review}.
Furthermore, SHSD has several important advantages over EDMD, in addition
to its simplicity:

\begin{enumerate}
\item SHSD has several controllable parameters that can be used to change
the transport coefficients and compressibility, notably the usual
density $\phi$ but also the cross-section factor $\chi$ and others%
\footnote{For example, one can combine rejection-free Maxwell collisions with
hard-sphere collisions in order to tune the viscosity without affecting
the compressibility. The efficiency is significantly enhanced when
the fraction of accepted collisions is increased, however, the compressibility
is also increased at a comparable collision rate.%
}, while EDMD only has density.
\item SHSD is time-driven rather than event-driven thus allowing for easy
parallelization.
\item SHSD can be more easily coupled to continuum hydrodynamic solvers,
just like ideal-gas DSMC \citet{FluctuatingHydro_AMAR} and DPD \citet{TripleScale_Rafael,DPD_IncompressibleNS}.
Strongly-structured particle systems, such as fluids with strong interparticle
repulsion (e.g., hard spheres), are more difficult to couple to hydrodynamic
solvers \citet{FluctuatingHydroHybrid_MD} than ideal fluids, such
as MPCD or (I-)DSMC, or weakly-structured fluids, such as DPD or SHSD
fluids.
\end{enumerate}
Finally, the stochastic particle model on which SHSD is based is intrisically
interesting and theoretical results for models of this type will be
helpful for the development of consistent particle methods for fluctuating
hydrodynamics.

\begin{acknowledgments}
The work of A. Donev was performed under the auspices of the U.S.
Department of Energy by Lawrence Livermore National Laboratory under
Contract DE-AC52-07NA27344 (LLNL-JRNL-415281). We thank Ard Louis
for sharing his expertise and code for solving the HNC equations for
penetrable spheres. We thank Salvatore Torquato, Frank Stillinger,
Andres Santos, and Jacek Polewczak for their assistance and advice.
\end{acknowledgments}

\end{document}